\documentclass[11pt]{article}
\usepackage[normalem]{ulem}	
\usepackage{graphicx}
\usepackage{amsmath}
\usepackage{amssymb}
\usepackage{amsthm, amscd}
\usepackage{amsfonts}
\usepackage{pstricks}
\usepackage{overpic}

\newcommand{\be}{\begin{equation}}
\newcommand{\ee}{\end{equation}}
\newcommand{\bea}{\begin{eqnarray}}
\newcommand{\eea}{\end{eqnarray}}

\newcommand{\note}[1]{$\color{red}{\bullet}$}

\begin{document}

\title{Exactly Solvable Lattice Models with Crossing Symmetry}
\author{Steven H.\ Simon$^1$ and Paul Fendley$^2$ \medskip\\
$^1$Rudolf Peierls Centre for Theoretical Physics, 1 Keble Road, Oxford, OX1 3NP, UK
\medskip\\
$^2$Department of Physics, University of Virginia,
Charlottesville, VA 22904-4714 USA
}

\smallskip

\date{December 13, 2012}

\maketitle
\begin{abstract}
We show how to compute the exact partition function for lattice statistical-mechanical models whose Boltzmann weights obey a special ``crossing'' symmetry.  The crossing symmetry equates partition functions on different trivalent graphs, allowing a transformation to a graph where the partition function is easily computed. The simplest example is counting the number of nets without ends on the honeycomb lattice, including a weight per branching. Other examples include an Ising model on the Kagom\'e lattice with three-spin interactions, dimers on any graph of corner-sharing triangles, and non-crossing loops on the honeycomb lattice, where multiple loops on each edge are allowed. We give several methods for obtaining models with this crossing symmetry, one utilizing discrete groups and another anyon fusion rules.  We also present results indicating that for models which deviate slightly from having crossing symmetry, a real-space decimation (renormalization-group-like) procedure restores the crossing symmetry.

\end{abstract}

\section{Introduction}

Over the decades, many techniques have been developed to ``solve'' classical lattice models., i.e.\ compute their partition function in the thermodynamic limit. These typically apply when the Boltzmann weights obey the Yang-Baxter equation \cite{Baxbook}. It has become clear that only models with special properties are amenable to such treatments; these models are usually called ``integrable''. Generic lattice models are certainly not integrable, but many of the simplest and most interesting models are, and much has been learned about physics as a consequence of these exact results.

In this paper we develop a very simple technique to compute exact partition functions of some classical lattice models. While only a very special class of models is amenable to this technique, such models do not seem to have been analyzed before; in general they do not appear to be special cases of known integrable models. The degrees of freedom of the models we study live on the {\em edges} of any lattice (or actually, graph) with trivalent vertices, such as the honeycomb lattice, or equivalently, the sites of the Kagom\'e lattice. In several of the simplest and most interesting cases, the models have nice geometric interpretations.

To give an example, the simplest case corresponds to counting the number of ``nets"  without ends on the honeycomb lattice. In this case there are two possible states on each edge, corresponding to covering the edge with a net, or leaving it empty. The counting problem becomes non-trivial by the requirement that  the net has no ends, i.e.\ no configurations are allowed where only one of the three edges touching a vertex is covered. We do allow all three edges touching a vertex to be covered, which corresponds to a ``branching''. A typical net configuration is illustrated in figure \ref{fig:typ} below. We show that the number of such net configurations on any planar graph with an even number $N$ of trivalent vertices is exactly
\begin{equation}
\left(\frac{5+\sqrt{5}}{2}\right)^{N/2} +\
\left(\frac{5-\sqrt{5}}{2}\right)^{N/2}\ .
\label{Zbp}
\end{equation}
This result can be extended to allow for an arbitrary weighting per branching, as we will explain in section \ref{sec:crossing}.

\begin{figure}[h]
 \begin{center}
 \includegraphics[scale=0.2, angle=90]{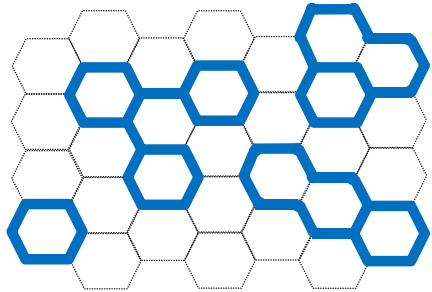}
\end{center}
 \caption{A typical configuration of closed nets.}
\label{fig:typ}
\end{figure}

Such computations are possible because in this special class of models, the Boltzmann weights obey a very special property we name ``crossing'' symmetry, by analogy with Feynman diagrams in field theory. Crossing symmetry allows us to show the partition functions on different lattices related by an ``F-move'' are identical. Repeating F-moves allows us to transform the lattice into one where the partition function is easily computed by diagonalizing an $M\times M$ matrix, where $M$ is the number of states allowed on each edge of the graph. This procedure of changing the lattice is reminiscent of techniques involving the Yang-Baxter equation (in particular that used to solve the eight-vertex model on the Kagom\'e lattice \cite{Baxbook}), but simpler and much less general. It is also reminiscent of numerical techniques used in some classical \cite{Loh,WuGuo} and quantum two-dimensional models \cite{LevinNave}, but our technique is not only exact, but can be easily implemented analytically.

The outline of this paper is as follows: In section \ref{sec:crossing}, we define crossing symmetry and F-moves for unoriented edges and vertices. We show how to solve models with this crossing symmetry,  illustrating this by nets without ends.   In section \ref{sec:loop} we show how certain loop gases have crossing symmetry, and we study some of the properties of these loop models.  In section \ref{sec:vertex} we solve more general ``vertex'' models, where both the edges and the vertices of the graph can be oriented.   One such example generalizes the counting of dimer configurations on a Kagom\'e lattice.   In section \ref{sec:moregeneral} we discuss several methods to generate models with
crossing symmetry.   In section \ref{sub:finite} edges are labeled with the elements of a finite group, whereas in section \ref{sub:anyon}, they are labeled with the elements of a quantum theory of anyons (i.e., the objects of a modular tensor category, or the primary fields of a rational conformal field theory).  In section \ref{sec:duality} we briefly discuss the relationship of these models based on anyon theories to the quantum string-net models of Levin and Wen \cite{LevinWen}.   Finally in section \ref{sec:decimation} we consider a real-space decimation transformation which is exact for models with crossing symmetry and we give some examples of these flows in section \ref{sub:someexact}.  In section \ref{sub:flowstocrossing} we consider models which are perturbed away from crossing symmetry and we show that under this decimation transformation they ``flow'' back to having crossing symmetry.  Finally, in section \ref{sec:summary} we summarize our results and discuss further possible directions. Several
appendices are included to give more details of certain calculations.

\section{Crossing symmetry and F-moves}
\label{sec:crossing}

The statistical-mechanical models we study are defined on any lattice or graph ${\cal G}$ with trivalent vertices, such as the honeycomb lattice.  For simplicity we restrict to those ${\cal G}$ that can be embedded on a sphere, but most of our results will apply to any graph of trivalent vertices.  

The degrees of freedom live on the {\em edges} of this lattice, and there are $M$ possible states labeled by an index $i \in 0\dots M-1$. To allow for more general ``vertex''-type models, one can orient each edge, but we defer this (only slightly more complicated) case to section \ref{sec:vertex} below. Here the Boltzmann weights depend only on the three states around each vertex, so we label them by $w(j,k,l)$.

\begin{pspicture}(-1,-0.3)(4,2.3)
\psline(1,1)(1,2)
\psline(1,2)(1,1.5)
\psline(0,0)(1,1)
\psline(0,0)(.5,.5)
\psline(2,0)(1,1)
\psline(2,0)(1.5,.5)
\put(1.1,1.5){$j$}
\put(.2,.4){$k$}
\put(1.7,.4){$l$}
\put(2.5,1){$=w(j,k,l) = w(k,l,j) = w(k,j,l)$.}
\end{pspicture}

\noindent For now we will further assume that $w(j,k,l)= w(k,j,l)$, but in section  \ref{sec:vertex} below we will study the more general case where $w$ depends on the order of its arguments.
The Boltzmann weight of each configuration is the product of all these vertex Boltzmann weights, and the partition function is then the sum over all possible configurations. For a model on ${\cal G}$
\be
 Z ({\cal G})= \sum_{\rm edge\  labels}\,  \prod_{{\rm vertices}\ v} w(j_v,k_v,l_v)
 \label{Zgen}
 \ee
where $j_v, k_v, l_v$ are the labels on the edges touching vertex $v$ in a given configuration. Note that we have defined the model so that the partition function depends only on the topology of ${\cal G}$ and the vertex weights $w$.

\begin{figure}[h]
 \begin{center}
 \begin{overpic}[scale=.5]{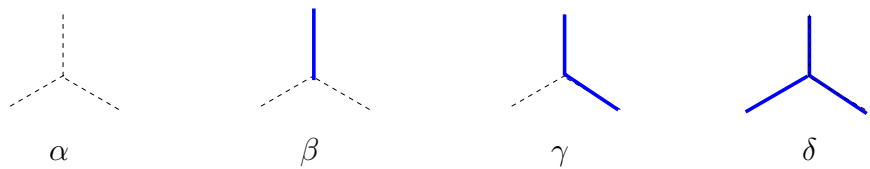}
\put(5,-5){\large $\alpha$}
\put(34,-5){\large $\beta$}
\put(63,-5){\large $\gamma$}
\put(92,-5){\large $\delta$}
 \end{overpic}
 \end{center}
 \vspace*{15pt}
 \caption{The four types of configurations for a model with $M=2$ types of edges.  For closed nets, there are no endpoints, so $\beta=0$, while nets have weight per unit length 1 ($\alpha=\gamma=1$) and branches have a weight $z$ ($\delta=z$) .}
 \label{fig:bp1}
\end{figure}

A simple example to keep in mind is that of  closed nets. Here $M=2$, with the two possible states on each edge corresponding to the net covering a edge or not. The four types of configurations around each vertex and their weights are displayed in figure \ref{fig:bp1}.  We use the word  ``closed'' to mean that net ends are forbidden, i.e.\ the Boltzmann weight $\beta=0$.   There are 8 possible configurations at each vertex, the four in the figure with
the other four found by rotating the two middle pictures. We will show that we can count the number of net configurations without ends, including a weight $z$ per branching. The corresponding Boltzmann weights are thus chosen to forbid ends ($\beta=0$),  to give a weight per unit length of $\gamma=\alpha=1$, and $\delta=z$.
The partition function with these Boltzmann weights then reduces to a sum over closed nets ${\cal C}$:
\be
Z_{\rm nets} = \sum_{\cal C} z^{\cal B}\ ,
\label{Znets}
\ee
where ${\cal B}$ is the number of branchings.  A typical closed-net configuration is shown in figure \ref{fig:typ}.  Note that, unlike the domain walls in an Ising or Potts model, barbell-type configurations are allowed (i.e., the covered edges need not separate two regions from each other).

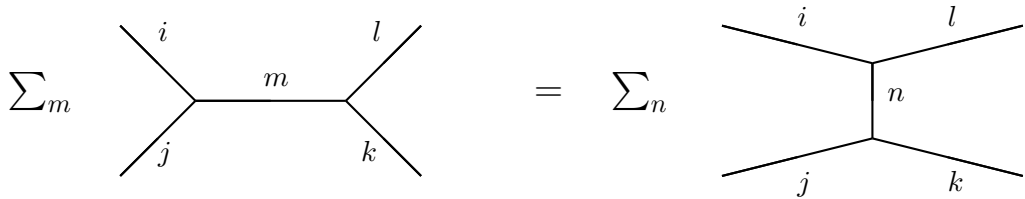
\begin{figure}[h]
\hspace*{40pt}\begin{pspicture}(-1,-1.4)(3,1.5)
\put(-.5,0){$\mbox{\Large $\sum_m$}$}

\psline(1,1)(2,0)
\psline(1,1)(1.5,.5)
\psline(1,-1)(2,0)
\psline(1,-1)(1.5,-.5)
\psline(2,0)(4,0)
\psline(2,0)(3,0)
\psline(4,0)(5,1)
\psline(5,1)(4.5,.5)
\psline(4,0)(5,-1)
\psline(5,-1)(4.5,-.5)
\put(1.5,.8){$i$}
\put(1.5,-.8){$j$}
\put(4.35,.8){$l$}
\put(4.2,-.8){$k$}
\put(2.9,.2){$m$}
\put(6.5,0){$\mbox{\Large $= \quad\sum_n$}$}

\psline(9,1)(11,.5)
\psline(9,1)(10,.75)
\psline(9,-1)(11,-.5)
\psline(9,-1)(10,-.75)
\psline(11,.5)(11,-.5)
\psline(11,.5)(11,0)
\psline(11,.5)(13,1)
\psline(13,1)(12,.75)
\psline(11,-.5)(13,-1)
\psline(13,-1)(12,-.75)
\put(10,1){$i$}
\put(10,-1.2){$j$}
\put(12,1){$l$}
\put(12,-1.2){$k$}
\put(11.2,0){$n$}
\end{pspicture}
\caption{The definition of crossing symmetry.}
\label{fig:firstcrossing}
\end{figure}

The models we solve are invariant under what we call a {\em crossing symmetry}. This is a relation between a product of the Boltzmann weights of two neighboring vertices
\be
 \sum_{m=0}^{M-1} w(i,j,m) w(m,k,l) = \sum_{n=0}^{M-1} w(l,i,n) w(n,j,k)
 \label{crossing}
\ee
as illustrated by Fig.~\ref{fig:firstcrossing}.

\noindent
For the model to be solvable by our method, this relation must hold for any choice of $i,j,k,l$.
It is simple to check that the Boltzmann weights displayed in figure \ref{fig:bp1} satisfy crossing symmetry for the case of closed nets with weighted branchings.  For this case, the only time the sum over the internal labels involves more than one term is when all four external legs are covered by nets.

In general, a useful way of expressing crossing symmetry comes from rewriting the weights $w(i,j,k)$ as a set of $M$ symmetric matrices $\Phi^{(i)}$ with elements $\Phi^{(i)}_{jk}=w(i,j,k)$. The crossing condition Eq.~\ref{crossing} then can be rewritten as
\be
\sum_{m} \Phi^{(i)}_{jm}\Phi^{(k)}_{ml} = \sum_{n} \Phi^{(k)}_{jn}\Phi^{(i)}_{nl}\ ,
\label{matrixcrossing}
\ee
where here and henceforth all sums run from $0$ to $M-1$. Crossing symmetry is therefore
equivalent to the statement that all of the matrices $\Phi^{(i)}$ commute.

The ``crossing'' name comes from analogy with field theory, where each trivalent vertex represents the fusion of two operators into a third under the operator product expansion. Thus multiplication by the matrix $\Phi^{(i)}$ can be thought of as fusing with a particle of type $i$, such that the matrix element $\Phi^{(i)}_{jk}$ means fusing a particle of type $i$ with with one of type $j$ to get a particle of type $k$. Crossing symmetry then means the four-point function can be decomposed into a sum of intermediate states using these operator products, with the result independent of the order in which the decomposition is done. While our notion of crossing symmetry is much less intricate than that in field theory, we do adopt another convenient piece of language from conformal or topological field theory, where such decompositions are quite useful: we call the process of exchanging the pictures on the left and right of  Fig.~\ref{fig:firstcrossing} an {\em F-move}.

Crossing symmetry in our context means that the partition function on graphs related by F-moves is identical. Namely when $u$ and $v$ are adjacent vertices, and $F_{uv}({\cal G})$ is the graph obtained from ${\cal G}$ by doing an F-move on $u$ and $v$, then
\[ Z(F_{uv}({\cal G})) = Z({\cal G})\ .\]
For this to be true, it is essential that all states be summed over: the equality is obviously not true configuration by configuration.

The equality of partition functions under crossing symmetry allows us to relate any planar graph of trivalent vertices to one where the partition function can be simply calculated. For example, consider the following series of F-moves:

\begin{pspicture}(-3,-1.5)(3,1.5)
\psline(1,1)(1,-1)
\psline(1,1)(-1,1)
\psline(-1,1)(-1,-1)
\psline(-1,-1)(1,-1)

\psline(.5,.5)(.5,-.5)
\psline(-.5,-.5)(.5,-.5)
\psline(-.5,-.5)(-.5,.5)
\psline(.5,.5)(-.5,.5)

\psline(.5,.5)(1,1)
\psline(.5,-.5)(1,-1)
\psline(-.5,-.5)(-1,-1)
\psline(-.5,.5)(-1,1)

\psline[arrowsize=10pt,linewidth=4pt]{->}(2,0)(3,0)

\psline(6,1)(6,-1)
\psline(4,-1)(6,-1)

\pscurve(4,-1)(4.1,-.5)(4.25,0)
\pscurve(4.5,-.5)(4.4,-.25)(4.25,0)

\psline(5.5,.5)(5.5,-.5)
\psline(4.5,-.5)(5.5,-.5)

\pscurve(6,1)(5.5,.9)(5,.75)
\pscurve(5.5,.5)(5.4,.6)(5,.75)

\pscurve(5,.75)(4.75,.5)(4.5,.25)(4.25,0)

\psline(5.5,.5)(6,1)
\psline(5.5,-.5)(6,-1)
\psline(4.5,-.5)(4,-1)

\psline[arrowsize=10pt,linewidth=4pt]{->}(7,0)(8,0)

\pscircle(9,0){.2}
\pscircle(9.5,.5){.2}
\pscircle(10,0){.2}
\pscircle(9.5,-.5){.2}

\psline(9,0.2)(9.3,.5)
\psline(10,0.2)(9.7,.5)
\psline(10,-0.2)(9.7,-.5)
\psline(9,-0.2)(9.3,-.5)

\end{pspicture}

\noindent
We applied the F-move once in the first step and thrice in the second step, reducing the graph to a one-dimensional chain of bubbles. Any planar graph of trivalent vertices can be similarly reduced to a chain of bubbles, because each individual face on the graph can be reduced by F-moves to be of bubble form. Repeating this for a graph with no dangling edges yields a one-dimensional  closed chain of bubbles. Since under an F-move the number of vertices in the graph remains the same, given $N$ vertices in the initial graph, using the crossing rule we can reduce it to a bubble chain containing $N/2$ bubbles.

In this form, the model can then be solved trivially by transfer matrix.  We write the weight for a single bubble, summed over internal lines, as

\hspace*{20pt} \begin{pspicture}(-2,0)(3,2)
\put(-1,1){$T_{ij} = \,\,\,\,\,\,\,\,\, \sum_{n,m} $}
\put(2.5,1.1){$i$}
\psline(2.5,1)(3,1)
\put(3.4,1.6){$n$}
\pscircle(3.5,1){.5}
\put(3.35,.2){$m$}
\psline(4,1)(4.5,1)
\put(4.3,1.2){$j$}
\put(6,1){$ = \,\,\,\,\,\,\, \sum_{n,m}  w(i, m,n)\, w(j, n, m) $}
\end{pspicture}

\noindent
The resulting transfer matrix with elements $T_{ij}$ is an  $M\times M$ matrix, in terms of which the partition function is
\begin{equation}
 Z({\cal G}) = {\rm Tr}[T^{N/2}]\ .
 \label{ZT}
 \end{equation}
 Since the partition function is invariant under F-moves, this applies to {\em any}  graph comprised of $N$ trivalent vertices. Thus computing the partition function of models with crossing symmetry requires only diagonalizing an $M\times M$ matrix!

Rewriting the weights in terms of the real symmetric matrices $\Phi^{(i)}$ as above, the crossing condition means that the matrices all commute. As a consequence, the $\Phi^{(i)}$ have the same eigenvectors and can be simultaneously diagonalized.
Thus we define real orthonormal eigenvectors $v^{(a)}$ and eigenvalues $\phi^{(i,a)}$ such that
$$
\sum_{k} \Phi^{(i)}_{jk} v^{(a)}_k = \phi^{(i,a)} v^{(a)}_j\ .
$$
In this language the transfer matrix can be written as $T_{ij}= {\rm Tr}[\Phi^{(i)}\Phi^{(j)}] = \sum_{n}  \phi^{(i,n)}\phi^{(j,n)}$, or as
\begin{equation}
  T_{ij} = 
 \sum_{m,n}  \Phi^{(n)}_{im} \Phi^{(n)}_{mj}
 = \sum_{a,n} v^{(a)}_i [ \phi^{(n,a)}]^2 \, v^{(a)}_j
\end{equation}
where we used $\sum_a v^{(a)}_l v^{(a)}_m =\delta_{lm}$.
The eigenvalues of $T$ are thus given by
\begin{equation}
 \lambda^{(a)} = \sum_{n}  [ \phi^{(n,a)}]^2
\label{eq:eigsT}
\end{equation}
so that the partition function is  given by
\be
\label{eq:myeq1}
  Z = \sum_ a\left(\sum_n  [\phi^{(n,a)}_{i} ]^2 \right)^{N/2}\ .
\ee

For closed nets as shown in Fig.~\ref{fig:bp1} and partition function (\ref{Znets}), the transfer matrix is
\be
T_{\rm nets} =
\begin{pmatrix}
2&z\\
z&2+ z^2
\end{pmatrix}\ ,
\ee
which has eigenvalues
$$
\lambda_\pm \equiv \frac{1}{2} \left( z^2+4 \pm z \sqrt{z^2+4}\right)\ .
$$
The partition function is thus
\begin{equation}
 Z_{\rm nets} = (\lambda_+)^{N/2}\ +\ (\lambda_-)^{N/2}
\label{Zeig}
\end{equation}
The number of such closed net configurations is given by setting $z=1$, yielding the result (\ref{Zbp}) quoted in the introduction. The free energy per vertex in the limit of a large number of vertices is $\ln(\lambda_+)/2$, which for $z=1$ is $\ln[2\cos(\pi/10)]$. This partition function allows the average number of branchings to be computed easily as
\[  \langle {\cal B} \rangle = z\frac{d}{dz} \ln(Z_{\rm nets})\ .\]

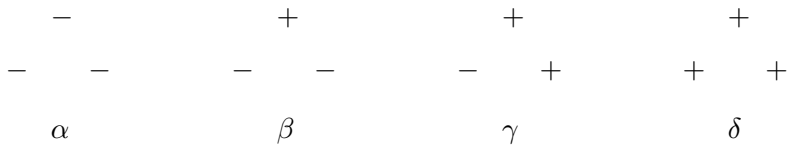
\begin{figure}[h]
\begin{pspicture}(-2,-0.3)(4,2.3)
\put(1,1.5){$-$}
\put(.4,.8){$-$}
\put(1.5,.8){$-$}
\put (1,0){$\alpha$}
\put(4,1.5){$+$}
\put(3.4,.8){$-$}
\put(4.5,.8){$-$}
\put (4,0){$\beta$}
\put(7,1.5){$+$}
\put(6.4,.8){$-$}
\put(7.5,.8){$+$}
\put (7,0){$\gamma$}
\put(10,1.5){$+$}
\put(9.4,.8){$+$}
\put(10.5,.8){$+$}
\put (10,0){$\delta$}
\end{pspicture}
\caption{The same weightings of the two state model as shown in Fig.~\ref{fig:bp1} but written in the language of an Ising model.}
\label{fig:weightings}
\end{figure}

A more general two-state model comes by relaxing the restriction that there be no net ends.
Such a model can also be naturally described as an Ising-type model on any graph ${\cal M}$ of {\em corner-sharing triangles}. ${\cal M}$ is the {\em medial graph} of ${\cal G}$: each vertex of ${\cal M}$ corresponds to an edge of ${\cal G}$, and such vertices are connected by an edge of ${\cal M}$ if the corresponding edges of ${\cal G}$ meet at a vertex of $\cal G$.  For ${\cal G}$ comprised of trivalent vertices,  ${\cal M}$  is comprised of corner-sharing triangles, so for example when ${\cal G}$ the honeycomb lattice, ${\cal M}$ is the Kagom\'e.  The degrees of freedom live on the sites of ${\cal M}$, so that the configurations in Fig.~\ref{fig:bp1} are rewritten as in Fig.~\ref{fig:weightings}, where $\alpha=w(---)$, $\beta=w(+--)$, etc.
\hspace*{80pt}

The crossing relation (\ref{crossing}) is not satisfied for arbitrary Boltzmann weights in this two-state model, but rather requires that the weights obey.
\be
\alpha \gamma +\beta\delta = \beta^2+\gamma^2
\label{twostate}
\ee
The special case of the closed net model ($\alpha=\gamma=1$, $\beta=0$, and $\delta=z$) indeed satisfies this crossing relation. The transfer matrix is then
\be
T= \left( \begin{array}{cc}  \alpha^2 + 2 \beta^2 + \gamma^2   &   \alpha\beta + 2 \beta\gamma + \gamma\delta  \\   \alpha\beta + 2 \beta\gamma + \gamma\delta & \beta^2 + 2 \gamma^2 +\delta^2 \end{array} \right)
\ee
The partition function is again expressed in the form (\ref{Zeig}), but the explicit form of the eigenvalues does not seem to be particularly illuminating here.

The solution (\ref{twostate}) of the crossing relations is the most general solution with $M=2$ states on each unoriented edge. We find it remarkable that of the three-parameter family of distinct Boltzmann weights (one weight can always be rescaled to one), a two-parameter family satisfies the crossing relation. One can work out the analogous solution for $M=3$ states on each edge; we give it explicitly in the appendix \ref{app:unoriented}. There are 10 distinct configurations around each vertex (up to rotations), and a three-parameter family of solutions to the crossing relations.
Obviously such general solutions will get even more unwieldy as the number of allowed states on each edge is increased further. However, various special cases can be quite elegant; in the next section, we discuss one of them.

One thing worth noting is that there is no way to make the two eigenvalues equal in the $M=2$ models with positive Boltzmann weights.  (The eigenvalues are equal only in the following three cases: $\beta=\gamma=0$ and $\alpha=\delta$;  $\alpha=\gamma=0$ and $\beta=\delta$; $\beta=\delta=0$ and $\alpha=\gamma$.)   There is no physically interesting phase transition as a function of the Boltzmann weights, a fact that remains true for all models solvable via crossing symmetry.  This is perhaps not surprising given that the partition function is the same as that of a one-dimensional system. A phase transition would occur when the two largest eigenvalues of the transfer matrix cross, as this gives a cusp in the free energy. If all the Boltzmann weights are non-negative, then all of the entries in the transfer matrix are non-negative, and by the Perron-Frobenius theorem we cannot have a degenerate largest eigenvalue.  If some Boltzmann weights are zero, the only way in which such a degeneracy can occur is if the transfer matrix decomposes into blocks. In the one-dimensional model, this type of (first-order) phase transition is somewhat trivial:  In one phase all of the incoming edge variables are from one set and in the other phase all of the incoming edge variables are from a completely disjoint set.  (For example, for the above mentioned case  of $\beta=\gamma=0$, the transition is between a system that is all spin up to a system that is all spin down).   However, in more complicated cases with more possible states on each edge, the nature of this phase transition is much less obvious.

\section{Loop models}
\label{sec:loop}

We showed in the previous section \ref{sec:crossing} that whenever the Boltzmann weights on a trivalent graph satisfy the crossing relation (\ref{crossing}), the partition function can be easily computed. Our next step is to understand better what sorts of models can be solved in this fashion. In this section we describe a general family of models having crossing symmetry that are naturally written in terms of loops.   In the next section we will introduce vertex models and present a several more general classes of solutions to the crossing relations.

\begin{figure}[h]
 \begin{center}
 \includegraphics[scale=0.5]{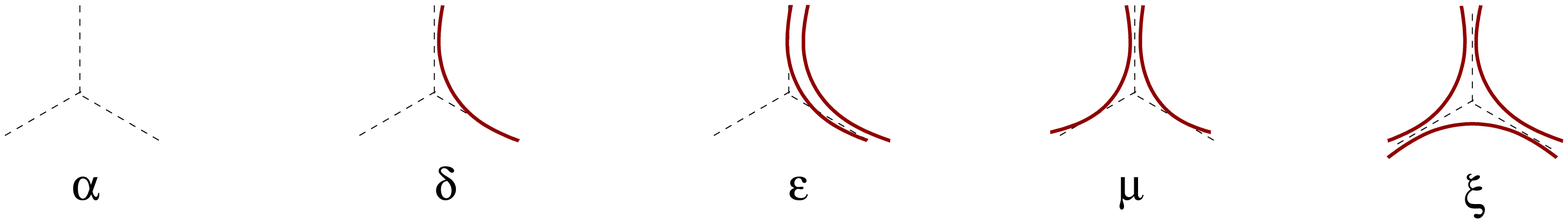}
\end{center}
 \caption{The five types of configurations and their weights in the three-state loop model. The middle three may be rotated, giving 11 possible configurations at each vertex. }
\label{fig:loopvertex}
\end{figure}

In this section we describe models of closed non-intersecting loops on ${\cal G}$.
Each of the $M$ states $i=0\dots M-1$ on each edge is pictured by drawing $i$ loops on each edge. The configurations are then required to be comprised of loops that {\em do not cross} at each vertex.  For $M=2$, this corresponds to setting $z=0$ in the closed net model, as apparent in figure \ref{fig:bp1}. Counting these $M=2$ loop configurations is trivial: the partition function is then simply $Z=2^{N/2+1}$, as can be seen directly by noting that the loops are domain walls for an Ising model on the dual graph. For $M=3$, however, the loop model is not trivial: the five types of vertices where loops do not cross are illustrated in figure \ref{fig:loopvertex}.

It follows from appendix \ref{app:unoriented} or an easy direct calculation that the weights of our $M=3$ loop model satisfy the crossing relation for $\alpha=\delta=\epsilon=1$ and $\xi=\mu-1/\mu$. There thus is a one-parameter family of solutions where the partition function can be computed exactly by reducing the graph to bubbles by F-moves. In this case the transfer matrix simplifies to
\[
T =
\begin{pmatrix}
3&0&2\mu-\mu^{-1}\\
0&2\mu^2+2&0\\
2\mu-\mu^{-1}&0&2\mu^2+\mu^{-2}
\end{pmatrix}
\]
which has eigenvalues $2(\mu^2+1)$, $2(\mu^2+1)$ and $1+\mu^{-2}$.  We will see below that this degeneracy of the largest eigenvalues, which indicates the system is tuned to a phase transition (for any value of $\mu$), is generic for loop models.

To systematically generalize this solution to higher $M$, it is useful to recall another context in which the crossing relation (\ref{crossing}) arises. This is as a consistency constraint for the {\em fusion rules} for operators in a theory of anyons (or equivalently, a rational conformal field theory or topological field theory). In such theories the vertex weight $w(j,k,l)$ is non-zero when one of the particles  is a ``bound state'' of the other two particles. The crossing constraint then guarantees that a ``singlet'' bound state of four particles can be found by considering any grouping pair by pair.

A simple case to illustrate this idea is if  $M\to\infty$ and each state $j$ labels the representation of the group or algebra $SU(2)$ with spin $j/2$. Then we let $w(j,k,l)=1$ if the tensor product of the representations of spin $j/2$ and spin $k/2$ contains the representation of spin $l/2$, and $w(j,k,l)=0$ otherwise. Thus $w(j,k,l)=1$ when $l\in |j-k|/2, |j-k|/2+1, \dots (j+k)/2$; this is commonly known as the triangle inequality for addition of angular momentum.  Note that this definition is symmetric in $j$, $k$ and $l$.   These weights do satisfy the crossing constraint (\ref{crossing}), which has a simple interpretation in the $SU(2)$ context: If one takes the tensor product of  four representations of spin $i/2$, $j/2$, $k/2$ and $l/2$ respectively, either side of the crossing equation corresponds to the number of ways this tensor product gives the trivial representation.  For example, if all four representations are spin 1/2, each side is 2:
\[(1/2)\otimes (1/2) \otimes (1/2) \otimes (1/2) = \{(0)+(1)\}\otimes \{(0) + (1)\} = (0) + (1) + (1) + (0) + (1) + (2) \ .\]
The crossing relation shows that to obtain the correct number, it does not matter how the representations are initially paired to do the tensor products.  The idea of using a fusion algebra to give solutions of the crossing equation turns out to be quite general. In fact, in section \ref{sub:anyon} we will explain how the fusion rules of any consistent anyonic theory can be utilized to find a solution of the crossing constraint.

Fusion provides a nice way of finding more general loop models solvable by our method. As above, group the vertex weights into a set of $M$ matrices $\Phi^{(i)}$, with entries $\Phi^{(i)}_{jk} = w(i,j,k)$.  Satisfying the crossing constraints is equivalent to demanding that the matrices $\Phi^{(j)}$ all commute with each other. For the three-state loop model satisfying the crossing relation, we have $\Phi^{(0)}$ the identity matrix, while
\be \Phi^{(1)}= \begin{pmatrix}
0&1&0\\
1&0&\mu\\
0&\mu&0
\end{pmatrix}
\ , \qquad
 \Phi^{(2)}= \begin{pmatrix}
0&0&1\\
0&\mu&0\\
1&0&\mu-1/\mu
\end{pmatrix}
\label{phi12}
\ee
These matrices satisfy
\[ \Phi^{(1)}\Phi^{(1)} =   \mu\Phi^{(2)}+ \Phi^{(0)},\qquad \Phi^{(1)}\Phi^{(2)} =   \mu \Phi^{(1)}\ . \]
Ignoring the factor of $\mu$, the first of these obviously resembles the fusion rules for $SU(2)$, with fusion of the state $j$ akin to taking the tensor product with the spin-$j/2$ representation  (i.e., resembling two spin 1/2's fusing to form either spin 0 or spin 1). The second also resembles the $SU(2)$ fusion rules, with the caveat that $\Phi^{(3)}$ vanishes for this case with $M=3$.

This suggests defining loop models with higher values of $M$ via analogous recursion relations. 
Let $\Phi^{(0)}=\alpha_0 I$ with $I$ the $M\times M$ identity matrix, and
\be
\Phi^{(1)} = \begin{pmatrix}
0&\alpha_0&0&0&\dots&0 &0&0\\
\alpha_0&0&\alpha_1&0&\dots &0&0&0 \\
0&\alpha_1&0&\alpha_2&&0&0&0\\
\vdots&&&&&&&\vdots\\
0&0&0&0&\dots &\alpha_{M-3}&0&\alpha_{M-2}\\
0&0&0&0&\dots&0&\alpha_{M-2}&0
\end{pmatrix}
\label{phi0matrix}
\ee
i.e.\ $\Phi^{(1)}_{kl} =  \alpha_{l}\delta_{k,l+1}+ \alpha_{k}\delta_{k,l-1} $
with $k,l=0\dots M-1$. This is the most general set of Boltzmann weights for fusing one line with $k$ lines to get either $k+1$ or $k-1$ lines, consistent with non-intersecting loops.
In the $M=3$ case above, $\alpha_0=1$ and $\alpha_1=\mu$.
Then define the remaining weights for fusing more than one line via the
the matrix recursion relation
\be
\Phi^{(j+1)}=\frac{1}{\alpha_{j}}\left(\Phi^{(1)}\Phi^{(j)} - \alpha_{j-1}\Phi^{(j-1)}\right)\ .
\label{recursion}
\ee

The Boltzmann weights defined by the recursion relation (\ref{recursion}) satisfy the crossing constraint because all of the matrices $\Phi^{(k)}$ commute with each other. This is easy to prove. Suppose $\Phi^{(i)}$ commutes with all $\Phi^{(j)}$ for $j < i$. Using the recursion relation it is easily shown that $\Phi^{(i+1)}$ also commutes with all $\Phi^{(j)}$ for $j < i+1$.  Since $\Phi^{(0)}$ is proportional to the identity, it commutes with all $\Phi^{(i)}$, so by induction all $\Phi^{(i)}$ commute. Moreover,  $w(j,k,l)=\Phi^{(j)}_{kl}$ as defined here is symmetric in all three indices; we prove this in the appendix \ref{app:symm}.

The $w(j,k,l)$ defined by this recursion relation therefore make sense as Boltzmann weights obeying the crossing condition. These loop models therefore can be solved by our method for any $M$.  In appendix \ref{app:pairs} we show that for even $M$ the eigenvalues of $T$ are always two-fold degenerate, whereas for odd $M$ eigenvalues come in pairs except for one remaining unpaired eigenvalue (which cannot be the largest eigenvalue of $T$).
As mentioned above, by the Perron-Frobenius theorem, such a degeneracy of eigenvalues only occurs if the transfer matrix is block diagonal.  Indeed, it is easy to see that this is precisely what happens --- the transfer matrix preserves the even or odd parity of the number of loops on an edge, and is hence block diagonal.

\begin{figure}[h]
 \begin{center}
 \includegraphics[scale=0.5]{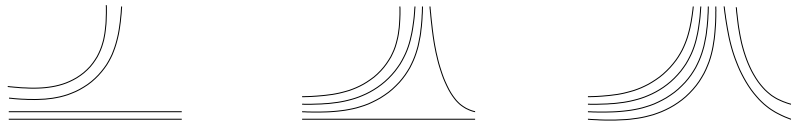}
\end{center}
 \caption{Illustrating how the recursion relation (\ref{recursion}) results in a non-intersecting loop model. When four lines enter from the left, and two enter from the right, the triangle inequality and conservation of parity (Eqs.~\ref{eq:tri1} and \ref{eq:tri2}) then require that the number of lines  going out the top must be $2,4$ or $6$.  }
\label{fig:rec1}
\end{figure}

This solvable model defined by (\ref{recursion}) can be represented as a non-intersecting loop model with the index $k=0,\dots, M-1$ of the edge representing $k$  lines running along each edge.  This correspondence is illustrated in Fig.~\ref{fig:rec1}. Here, when indices $j$ and $k$ come into two edges of the vertex, a non-vanishing weight occurs only if the third vertex has index $l$ with
\begin{eqnarray}
& & ~~~~|j-k| \leq l \leq j+k \label{eq:tri1} \\
& & \mod(j+k+l,2)=0 \ .
\label{eq:tri2}
\end{eqnarray}
as required by the above described triangle inequality. 
Since these two relations hold for $k=1$ from (\ref{phi0matrix}), induction using (\ref{recursion}) shows that they remain true for all $k$.  By drawing pictures analogous to those in Fig.~\ref{fig:rec1}, it is easy to see that these two relations are the requirements for defining a non-intersecting loop model, with the $M-1$ loops the maximum allowed on each edge.

If one sets all the $\alpha_j=1$, all the Boltzmann weights are either $0$ or $1$. These fusion rules are those of the quantum-group algebra $U_q(sl_2)$, or the conformal field theory $SU(2)_{M-1}$. The  ``truncation'' of allowing only $M$ loops on each edge is familiar in these contexts. We will explain in more detail in section \ref{sub:anyon} how such cases are related to anyon fusion rules.

\section{Vertex models}
\label{sec:vertex}

In this section we generalize our procedure to solve more general ``vertex'' models. We still have $M$ possible states on each edge of any planar graph with trivalent vertices. Now, however, we orient each edge by placing an arrow on it. In vertex models, there is a conjugation operation given by a permutation $P$ of $[1,\ldots, M]$ of order two (i.e., $P^2=1$).  We denote the conjugate of an index $i$ as $\overline i$ , and the conjugation is implemented by reversing the orientation of the edge:

\hspace*{80pt}\begin{pspicture}(0,0)(4,1.5)
\psline[arrowsize=5pt]{->}(1,.5)(1.5,.5)
\psline(1.5,.5)(1.75,.5)
\put(1.25,.75){$i$}
\put(2.5,.5){$=$}
\psline(3.75,.5)(4,.5)
\psline[arrowsize=5pt]{->}(4.5,.5)(4,.5)
\put(4,.75){$\overline i$}
\end{pspicture}

\noindent
If all states are self-conjugate (i.e.\ $P$ is the identity), then we can omit all arrows and the model reduces to the unoriented case treated above. The Boltzmann weights of these lattice models depend on the three states around each vertex, here taken in counterclockwise order. We define a weight function $w(i,j,k)=w(j,k,i) = w(k,i,j)$ corresponding to a trivalent vertex where incoming edges have labels $i$,$j$, and $k$.

\hspace*{30pt}\begin{pspicture}(0,-.5)(4,2.5)
\psline(1,1)(1,2)
\psline[arrowsize=5pt]{->}(1,2)(1,1.5)
\psline(0,0)(1,1)
\psline[arrowsize=5pt]{->}(0,0)(.5,.5)
\psline(2,0)(1,1)
\psline[arrowsize=5pt]{->}(2,0)(1.5,.5)
\put(1.2,1.5){$i$}
\put(-.2,.2){$j$}
\put(2,.2){$k$}
\put(3,1){$=w(i,j,k) = w(j,k,i) = w(k,i,j)$}
\end{pspicture}

\noindent The partition function is still defined by (\ref{Zgen}) as before, now being careful that in the weights $w(i,j,k)$ the edge variables must always be listed in counterclockwise order.

When we can define an orientation for each vertex (for example if the graph is embedded into an orientable surface), then it is possible to have $w(i,j,k)\ne w(j,i,k)$. This is also possible for unoriented edges.   For $M=2,3$ however, crossing symmetry requires $w(i,j,k)=w(j,i,k)$; see appendix \ref{app:m3}.  In the remainder of the current section we consider only cases where $w(i,j,k)= w(j,i,k)$, but in section \ref{sub:finite} we show how models based on finite groups with larger $M$
do allow weights to depend on the orientation of the vertex.

Crossing symmetry for these oriented vertex models is similar to the unoriented case. Pictorially,

\begin{pspicture}(-1,-1)(3,1.5)
\put(-.5,0){$\mbox{\Large $\sum_m$}$}

\psline(1,1)(2,0)
\psline[arrowsize=5pt]{->}(1,1)(1.5,.5)
\psline(1,-1)(2,0)
\psline[arrowsize=5pt]{->}(1,-1)(1.5,-.5)
\psline(2,0)(4,0)
\psline[arrowsize=5pt]{->}(2,0)(3,0)
\psline(4,0)(5,1)
\psline[arrowsize=5pt]{->}(5,1)(4.5,.5)
\psline(4,0)(5,-1)
\psline[arrowsize=5pt]{->}(5,-1)(4.5,-.5)
\put(1.5,.8){$i$}
\put(1.5,-.8){$j$}

\put(4.35,.8){$l$}
\put(4.2,-.8){$k$}
\put(2.9,.2){$m$}

\put(6.5,0){$\mbox{\Large $= \sum_n$}$}

\psline(9,1)(11,.5)
\psline[arrowsize=5pt]{->}(9,1)(10,.75)
\psline(9,-1)(11,-.5)
\psline[arrowsize=5pt]{->}(9,-1)(10,-.75)
\psline(11,.5)(11,-.5)
\psline[arrowsize=5pt]{->}(11,.5)(11,0)
\psline(11,.5)(13,1)
\psline[arrowsize=5pt]{->}(13,1)(12,.75)
\psline(11,-.5)(13,-1)
\psline[arrowsize=5pt]{->}(13,-1)(12,-.75)
\put(10.2,1){$i$}
\put(10.2,-1){$j$}
\put(12,1){$l$}
\put(12,-1.2){$k$}
\put(11.2,0){$n$}
\end{pspicture}
\medskip

\noindent
i.e. in the most general case,  the crossing constraint becomes
\begin{equation}
\label{eq:crossingvertex}
 \sum_m w(\overline m,i,j) w(k,l,m) = \sum_n w(l,i,\overline n) w(j,k,n)\ .
\end{equation}
The procedure of reducing each graph to a chain of bubble graphs via F-moves is the same as in the unoriented case, and the transfer matrix becomes

\begin{pspicture}(-1,0)(3,2)
\put(-.5,1){$T_{ij} = \sum_{n,m} $}
\put(2.5,1.1){$i$}
\psline(2.5,1)(3,1)
\psline[arrowsize=7pt]{->}(2.5,1)(2.9,1)
\put(3.5,1.6){$n$}
\pscircle(3.5,1){.5}
\psline[arrowsize=6pt]{->}(3.55,1.5)(3.58,1.5)
\psline[arrowsize=6pt]{->}(3.55,.5)(3.58,.5)

\put(3.5,.3){$m$}
\psline(4,1)(4.5,1)
\psline[arrowsize=5pt]{->}(4,1)(4.4,1)
\put(4.5,1.1){$j$}
\put(6,1){$ = \sum_{n,m}  w(m,\overline j,n) w(i, \overline m, \overline n) $\ .}

\end{pspicture}

\noindent As in the unoriented case, the partition function $ Z({\cal G}) = {\rm Tr}[T^{N/2}]$ can be rewritten in terms of the eigenvalues of the fusion matrix.

In the simplest vertex model, $M=2$ and the two states are conjugate to each other. This is the only $M=2$ vertex model; if the states are self-conjugate the model reduces to the unoriented case considered in section \ref{sec:crossing}. Configurations here can be labeled by putting an arrow on each edge, so that the weight of each vertex depends on how many arrows are pointing in and out.
\begin{align}
\nonumber
 w({\rm in , in, in})= \alpha ,\qquad
& w({\rm out, out , out}) = \widetilde{\alpha},  \\
  w({\rm in, in , out}) = \beta,\qquad
&  w(\rm{out, out , in}) = \widetilde{\beta}\ .
\end{align}
Since every edge appears in two vertex weights, once as in and the other as out, these weights can be rescaled as
\begin{equation}\alpha\to \lambda^3\alpha,\ ~~~ \widetilde{\alpha}\to\ \lambda^{-3} \widetilde{\alpha},\ ~~~
\beta\to \lambda\beta,\ ~~~ \widetilde{\beta}\to\ \lambda^{-1} \widetilde{\beta}
\label{eq:rescale}
 \end{equation}
 for any $\lambda$ without changing the partition function.  The nontrivial crossing constraint is
$$
 \alpha\widetilde{\alpha} = \beta \widetilde{\beta}\ .
$$
The transfer matrix is
$$
T= \left( \begin{array}{cc}  3 \beta \widetilde{\beta}+\alpha\widetilde{\alpha}  &   2 \alpha \widetilde{\beta} + 2 \beta^2 \\  2 \widetilde{\alpha} \beta+ 2 \widetilde{\beta}^2&
3 \beta \widetilde{\beta}+\alpha\widetilde{\alpha}   \end{array} \right)\ ,
$$
which using the crossing condition has eigenvalues $4\beta\widetilde{\beta} \pm 2(\sqrt{\alpha\widetilde{\beta}^3}
+\sqrt{\widetilde{\alpha}{\beta}^3})$.

A simple but non-trivial special case is $\widetilde{\beta}=\alpha = 0$.   In this case there are four non-trivial vertex configurations --- only all out, or two ins and one out (in three rotations) are allowed.  This case corresponds to counting close-packed dimers on the {\em medial graph} ${\cal M}$, comprised of corner-sharing triangles as described in section \ref{sec:crossing}.  The vertex configuration in our model with all arrows out corresponds to having no dimers on the corresponding triangle, while a vertex with only arrow one out corresponds to having a dimer on the edge of the medial graph connecting two in-pointing edges of the original graph \cite{ElserZeng}. Setting $\widetilde{\alpha}=1$ then
means that $\beta$ is the
weight per dimer. Using our formalism then gives $\pm 2 \beta^{3/2}$ for the eigenvalues of the transfer matrix, so that the partition function is
\be
Z_{\rm dimer} =
\begin{cases}
2(2\beta^{3/2})^{N/2} &N/2 \hbox{ even}\\
0 & N/2 \hbox{ odd}
\end{cases}
\label{Zdimer}
\ee
This was derived for Kagom\'e directly from a transfer matrix approach long ago \cite{Elser}. Here $N$ is the number of vertices of the honeycomb graph or triangles of the Kagom\'e.

The $M=3$ vertex model has one state its own conjugate ($\overline{0}=0$) with the other two conjugates of each other ($\overline{1}=2$). Thus each configuration can be pictured by arrows for $1$ and $\overline{1}$, and empty links for $0$. In appendix \ref{app:m3}, we show that crossing symmetry does not allow for the weights here to depend on the orientation of the vertex. This allows for 10 distinct Boltzmann weights at each vertex (up to rotations), labeled as in the unoriented case. The general solution for the $M=3$ vertex model can be found in the appendix \ref{app:vertex}, again giving six conditions.

A variety of special examples of the crossing-symmetric $M=3$ vertex model can be found by demanding that some weights be zero, or that they obey some additional symmetry. An example of the former is the analog of the ``no net end'' condition  $\beta=\gamma=0$, for which one subsequent solution is  (in the notation of appendix \ref{app:m3})
\[\alpha=2\phi,\qquad \delta=\phi^2/\epsilon, \qquad \eta=\phi^2 \kappa/\epsilon^2,\qquad \mu=\phi^2\xi/\epsilon^2\ .\]
An example of the latter comes from demanding that the weights be invariant under ``charge-conjugation'' symmetry, i.e.
\begin{equation}
w(i,j,k) = w(\overline{i},\overline{j},\overline{k}) \end{equation}
so that here $\beta=\gamma$, $\delta=\epsilon$, $\eta=\xi$, and $\kappa=\mu$. The crossing constraint then results in two possible solutions for the remaining weights in terms of $\alpha$, $\beta$ and $\delta$:
\[\phi=\delta,\qquad \kappa=\xi= \frac{1}{2\beta}\left(\beta^2+2\delta^2-\alpha\delta\right)\ ,\]
or
\[ \phi= \alpha - \delta +\beta^2/\delta,\qquad \kappa-\xi= \beta(1-\phi/\delta),\qquad \kappa+\xi=\delta(\delta+\phi)/\beta\ .\]

\section{More general solutions }
\label{sec:moregeneral}

The purpose of this section is to find additional examples of statistical models solvable via crossing symmetry.   We will pursue two routes, one based on using finite groups, and the other using anyon theories/modular tensor categories. We also comment on a connection between these models and solvable {\em quantum} two-dimensional lattice models in section \ref{sec:duality}, but do not attempt a complete classification of models solvable by this method.

\subsection{Models based on finite groups}
\label{sub:finite}

Here we construct a vertex model for any finite group $G$, where each directed edge is labeled with group elements  $i,j,k, \ldots \in G$. Reversing an arrow on an edge corresponds with taking the inverse of a group element, so that the conjugate $\overline i = i^{-1}$.   We will show that this model obeys crossing symmetry when the Boltzmann weights for a vertex are given by
\begin{equation}
  w(i,j,k) = W[ {\rm{conjugacy \, class}}( i j k)]\ .
 \end{equation}
In other words, if the elements entering the vertex (in counterclockwise order) are $i,j,k$, one multiplies these elements together to form the group element $g=ijk$. Two group elements $g,h$ are in the same conjugacy class if there exists some element $q$ in the group such that $g=qhq^{-1}$.  Each conjugacy class can then be assigned an arbitrary non-negative Boltzmann weight, which we have here denoted by $W$.   Since $ijk$ is in the same conjugacy class as $kij$ we must have $w(i,j,k) = w(j,k,i) = w(k,i,j)$ as required.   We note in passing that models defined this way bear a resemblance to the recent work of Ref.~\cite{Henley}.

For an abelian group, each element is its own conjugacy class, so we can take an independent weight for each group element. It is easy to prove that this satisfies the crossing symmetry (\ref{eq:crossingvertex}), which here amounts to
\[ \sum_m W(ijm) W(m^{-1}kl) = \sum_n W(iln)W(n^{-1}jk)\ .\]
Letting $g_1=ijm$, $g_2=m^{-1}kl$, $h_1=iln$, and $h_2=n^{-1}jk$, we have $g_1g_2 = h_1 h_2\equiv x$. Then the crossing relation is equivalent  to
\[ \sum_{g_1} W(g_1) W(x(g_1)^{-1}) =  \sum_{h_1} W(h_1) W(x(h_1)^{-1}). \]
These are obviously the same for any $x$.

For a nonabelian group, $ijk$ in general is not equal to $jik$, so $w(i,j,k) \neq w(j,i,k)$. For this to make sense we must require all the vertices as well as the edges of the graph ${\cal G}$ to be given an orientation.
The proof of crossing symmetry here is not difficult but requires a bit of elementary group theory \cite{Hamermesh}. The character $\chi_R(g)$ of a group element $g$ is simply the trace of the matrix $D^{(R)}(g)$ corresponding to $g$ in the representation $R$. The character depends only on the conjugacy class of $g$, and any function of a conjugacy class can be expanded in terms of characters $\chi_R(g)$ of the irreducible representations $R$. We can thus write
$$
 w(i,j,k) = \sum_R \, A_R \, \chi_R(ijk)
$$
where the $A_R$ are arbitrary coefficients, and the sum is over all irreducible representations.   The required crossing condition, Eq.~\ref{eq:crossingvertex}, can then be rewritten as
\begin{equation}
\label{eq:tocollapse}
  \sum_m \sum_{R,R'} A_R A_{R'} \, \chi_R(klm)\chi_{R'}(m^{-1} i j) = \sum_n \sum_{R,R'} A_R A_{R'} \, \chi_R(jkn)\chi_{R'}(n^{-1} l i)
\end{equation}
Note that $\chi_R(a b) = \chi_R(b a)$.   The following lemma
\begin{equation}
\label{eq:lemma}
   \sum_b  \chi_R(ab) \chi_{R'}(b^{-1} c) = \frac{|G|}{n_R} \delta_{RR'} \chi_R(ac)
\end{equation}
proves that the two sides of (\ref{eq:tocollapse}) must be equal to each other, thus showing that these models always satisfy crossing symmetry.   Here $n_R$ is the dimension of representation $R$,  $|G|$ is the order of the group $G$, and the sum over $b$ is over all elements of the group.

This key lemma is proved from the following matrix manipulations:
\begin{eqnarray} \label{eq:ar1}
 \sum_b D_{\alpha \beta}^{(R)}(ab) D_{\gamma\delta}^{(R')}(b^{-1} c) &=&  \sum_{b,x,y} D_{\alpha x}^{(R)}(a)  D_{x \beta}^{(R)}(b) D_{\gamma y}^{(R')}(b^{-1}) D_{y \delta}^{(R')}(c) \\ \label{eq:ar2}
&=& \frac{|G|}{n_R} \delta_{RR'} \delta_{\beta\gamma}\sum_{x} D_{\alpha x}^{(R)}(a)  D_{x \delta}^{(R')}(c) \\
&=& \frac{|G|}{n_R} \delta_{RR'} \delta_{\beta\gamma} D_{\alpha \delta}^{(R)}(ac)
\end{eqnarray}
where the sum on $b$ is over all group elements, and the sums on $x,y$ are over matrix indices.  Here we have used the fact that $D$ is a representation so that $D(ab) = D(a) D(b)$ as matrices, and in going from (\ref{eq:ar1}) to (\ref{eq:ar2}) we have used the well known orthogonality theorem \cite{Hamermesh}
$$
 \sum_{b}  D_{x \beta}^{(R)}(b) D_{\gamma y}^{(R')}(b^{-1})  = \frac{|G|}{n_R} \delta_{RR'} \delta_{\beta\gamma}\delta_{xy}
$$
Finally using the fact that the character $\chi_R(g) = {\rm Tr} D^{(R)}(g)$, we take the appropriate traces of both  sides of (\ref{eq:ar1}) to obtain the desired lemma (\ref{eq:lemma}).

\subsection{Models based on anyon fusion rules}
\label{sub:anyon}

In section \ref{sec:loop} we found a general class of solutions to the crossing constraints in the unoriented case, giving us the sequence of loop models. We noted that these were (when all $\alpha_j=1$) the fusion rules for a theory of anyons. The mathematical structure behind such fusion rules is called a {\em modular tensor category}, and is familiar in physics not only in anyon theories, but also in rational conformal field theory (RCFT) and topological field theory. In this language, the net models with $z=0$ correspond to  ``Fibonacci'' fusion rules, whereas the loop models are generally called $SU(2)_{M-1}$ fusion rules.
In this section we explain how in general any such fusion rules can be used to find solutions of the crossing relation. The resulting solutions do not have any free couplings as in the loop models above; all weights are non-negative integers. However, as with our earlier analysis, one can then add and/or perturb couplings and find still more general solutions to the crossing relation.

An anyon is a particle in a quantum theory in two spatial dimensions whose statistics generalize bosonic and fermionic ones. When an abelian anyon is rotated by $2\pi$ around another, the wave function changes by a phase. In the non-abelian case, the wavefunction is expressed not as a complex scalar, but as a complex vector, describing a state within a degenerate space. Exchange of particles can cause a change of state within this space.

An anyon theory can generally be described by a set of  $M$ fundamental particle types, labeled by $i=0\dots M-1$  with $0$ being the trivial particle (in the corresponding RCFT, these are called primary fields/operators, with $0$ the identity operator). A pair of anyons behaves as an anyon as well, and the fusion rules describe how the statistics of each pair is described in terms of the fundamental types. Precisely, the fusion rules are described by a non-negative integer $N_{ij\overline k}=N^{k}_{ij}$, which gives how many times the anyon $k$ appears in the fusion of particles of type $i$ and $j$, labeled schematically  as
\[ (i)\times (j) = \sum_k N^{k}_{ij}\ (k).\]
In the RCFT context, a non-vanishing $N_{ij}^{k}$ says that the field $k$ appears in the operator product expansion of $i$ and $j$.
Studying such fusion algebras is a major endeavor;  reviews of the rules can be found in \cite{MS} in the RCFT context, and in appendix E of \cite{Kithoneycomb} in the anyon context. A simpler overview can be found in \cite{Preskill}. Various examples and a start of a classification scheme can be found in \cite{Rowell}.

Any consistent anyon theory/RCFT/modular tensor category defines a classical lattice model with crossing symmetry. Each edge label corresponds to an anyon type, and the Boltzmann weight $w(i,j,k)=N_{ijk}$. The fusion rules are symmetric
under permutation of particle types: $N_{ijk}$ is independent of the ordering of the indices. Because fusion is associative, such weights automatically obey the crossing constraint
\begin{equation}
\label{eq:fusionassociative}
\sum_m N^{i}_{j{\overline m}}N^k_{ml} = \sum_n N^{i}_{l\overline{n}} N^k_{nj}\ .
\end{equation}
By our procedure, this gives a solvable model.

Anyon theories (modular tensor categories), require a great deal of mathematical structure that is unnecessary for defining a classical statistical model with crossing symmetry.  For example, one must have a consistent rule for what happens when two anyons are moved around each other (i.e., braided) in two dimensions.  A slightly more general mathematical structure which does not define such a braiding is known as a {\it unitary fusion category} \cite{ZWang}.   The objects in such a category also obey fusion rules and have the same associativity condition (\ref{eq:fusionassociative}). It therefore can be used to define a classical statistical model with crossing symmetry, even though in general the objects of a unitary fusion category cannot be thought of as being actual particles.  Even the unitary fusion categories have a great deal of mathematical structure that is not necessary for defining a statistical model with crossing symmetry (for example, all of the weights are integers). It is an
interesting
open problem to fully classify the most general algebras that would have the necessary crossing symmetry.

Several well-known examples of fusion rules of anyon models underlie the statistical models described above. The Fibonacci fusion rule describe two types of self-conjugate anyons, which we label $I$ and $X$.  The ``identity'' particle $I$ fuses according to its name:  $I\times I= I$  and $I\times X= X$, while the anyon $X$ fuses as $X^2=I+X$. The resulting non-zero weights come from $N_{111}=N_{1XX}=N_{XXX}=1$, with other $N_{ijk}$  not related by permutations vanishing. This gives the net model with $z=1$; the identity particle corresponds to leaving the edge empty, while the $X$ particle corresponds to covering it. So-called Ising fusion rules describe three self-conjugate particles.  The particles are the identity $I$, and two additional particles $X$ and $Y$ (often known as $\sigma$ and $\psi$) fusing as $X^2=I+Y$, $XY=X$ and $Y^2=I$ with $N_{YYY}$ vanishing. This is precisely the loop model with $M=3$ and $\alpha_0=\alpha_1=1$, identifying $X$ and $Y$ with edges having one and two loops
respectively.
Note that with $\alpha_1=\mu=1$, the weight for a vertex with three double lines meeting indeed vanishes, as apparent from (\ref{phi12}). For the loop models with general $M$ and all  $\alpha_j=1$, all $N_{ijk}$ are zero or $1$, and these correspond to $SU(2)_{M-1}$ fusion rules.
(Note that Ising and $SU(2)_2$ modular tensor categories are slightly different, but they have identical fusion rules.)

Yet another three-state unoriented (self-conjugate) example is given by the fusion rules \cite{Rowell}
\begin{equation}
X^2= I+ Y,\qquad XY= X + Y,\qquad Y^2= I + X +Y\ .
\label{fusionex}
\end{equation}
This case, with $N_{XXI}=N_{YYI}=N_{XXY}=N_{XYY}=N_{YYY}=1$ and others besides permutations vanishing,
corresponds to having nets with labels $X$ and $Y$ on each strand such that $XXX$ vertices are forbidden.
As with the loops, we can allow for more general weights. Using the notation of the appendix \ref{app:m3} for the Boltzmann weights, the corresponding matrices of weights are
\[ \Phi^{(I)}=\begin{pmatrix}
\alpha&0&0\\
0&\delta&0\\
0&0&\epsilon
\end{pmatrix}\ ,
\qquad
\Phi^{(X)}= \begin{pmatrix}
0&\delta&0\\
\delta&\eta&\mu\\
0&\mu&\kappa
\end{pmatrix}
\ , \qquad
 \Phi^{(Y)}= \begin{pmatrix}
0&0&\epsilon\\
0&\mu&\kappa\\
\epsilon&\kappa&\xi
\end{pmatrix}
\]
where we have kept $\beta=\gamma=\phi=0$ to resemble the fusion rules (\ref{fusionex}). These matrices commute for more general couplings than the fusion rules, namely
\be \alpha=\delta=\epsilon=1,\qquad \mu^2+\kappa^2-\mu\xi-\eta\kappa=1\ ,\ee
giving one constraint in addition to having $\Phi^{(I)}$ remain the identity matrix.
With these conditions,
\[ \Phi^{(X)}\Phi^{(X)} =   \mu\Phi^{(Y)}+ \eta\Phi^{(X)}+\Phi^{(I)},\quad \Phi^{(X)}\Phi^{(Y)} =   \mu \Phi^{(X)}+\kappa\Phi^{(Y)} , \quad \Phi^{(Y)}\Phi^{(Y)} =   \xi\Phi^{(Y)}+ \kappa\Phi^{(X)}+\Phi^{(I)}. \]
further generalizing the loop recursion relation.  It seems very likely that any anyon theory can be deformed to generate a larger family of weights that satisfy crossing symmetry.

\subsection{Duality and the relationship to quantum string-net models}
\label{sec:duality}

The models we describe have a great deal in common with the Levin-Wen quantum-mechanical ``string-net" lattice models \cite{LevinWen}.    These string-net models were cleverly constructed to give lattice realizations of certain topological quantum field theories.  We will quote some of the results of this work and we refer the reader to the original paper \cite{LevinWen} for further details; see also \cite{Burnell} for an alternative physical viewpoint.

As with our statistical models, the Levin-Wen models assign a label $i\in 0\dots M-1$ to each edge of a trivalent graph.  We may choose these labels from an anyon theory (or any unitary tensor category) or from a finite group; in the latter case, the resulting Levin-Wen model become in essence a lattice gauge theory.  There is a one-to-one mapping between edge-label configurations in our lattice model and orthonormal basis vectors of the Hilbert space in the string-net model.   The precise connection between the two occurs when our statistical model consists of vertex weights $w$ which are all either zero or one, so that a configuration is either ``allowed" (meaning it has weight one) or ``disallowed" (meaning it has weight zero). In the quantum string-net model, the disallowed configurations correspond to a so-called vertex excitation of the model.  The Hamiltonian of the model can be tuned to give such excitations very high energy and thereby such excitations
can effectively be eliminated from (low-energy) consideration.  We have no need to write the actual Hamiltonian, luckily, as it is quite complicated.

Thus we now consider the quantum string-net models in the absence of any vertex excitations.   The classical partition function then is the dimension of the remaining low-energy Hilbert space, i.e.\  the number of configurations of the edge variables with only allowed vertices.  In the string-net model this space can be described in terms of the ground state and all possible excitations which do not involve the vertex excitations.   The non-vertex excitations are all associated with the faces of the model.   In particular, each face can can be labeled with a quantum number chosen from the same set of $M$ labels (the same labels as we had put on the edges), which in some abstract sense can be thought of describing a ``magnetic flux" through the face.  Moreover, each flux acts like an anyon and may fuse with others in different ways, akin to spins adding together to form either singlet or triplet representations.  Counting of the different ways they
may fuse together allows
one to determine the dimension of the low-energy Hilbert space in another way.

\begin{figure}[h]
 \begin{center}
 \includegraphics[scale=0.25,angle=270]{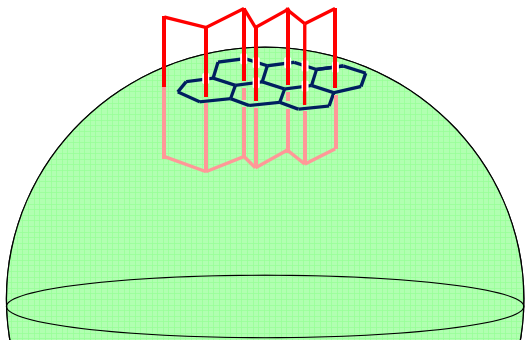}
\end{center}
 \caption{The graph ${\cal G}$ has seven black faces on the surface of the sphere (six hexagonal, and one very large face encompassing the remainder of the sphere).  The dual lattice has an edge (red) drawn through each face and hence through the sphere surface.  All of the endpoints inside and outside of the sphere are connected together using the minimal number of trivalent vertices.  Note that the number of vertices in the lattice (black) and the dual lattice (red) is identical (ten in each case). }
\label{fig:duality}
\end{figure}

In order to describe the connection between our statistical model and the quantum string-net models more clearly, it is useful to first show that these statistical models have an interesting geometric duality.   Consider a trivalent graph, and for simplicity assume that it is a planar graph embedded on a sphere.    Now construct a dual graph in the following way: draw an edge through each face normal to the sphere.   Using a minimal number of trivalent vertices, connect together all of the endpoints outside of the sphere and connect together all of the endpoints inside of the sphere.  Such a construction is shown explicitly in figure \ref{fig:duality}. Using the Euler characteristic on a sphere (${\rm{Faces}} + {\rm{Vertices}} - {\rm{Edges}} =2$), and the fact that all vertices are trivalent so (${\rm{Edges}} = (3/2){\rm{Faces}}$), it is easy to establish that the number of vertices in the dual graph is the same as the number of vertices in the original graph.  Thus the partition function of the statistical model on the dual graph is identical to the partition function on the original graph.

In the string-net model this dual graph has a very natural interpretation.  Each edge through a face is labeled with one of the $M$ possible labels, thus describing the flux through each face.  These fluxes can fuse together in all possible ways on both the inside and the outside of the sphere, and all of these possible so-called ``fusion channels" describe the full low-energy Hilbert space of the string-net model without any vertex excitations.  This is equivalent to counting the number of allowed edge-label configurations on the original graph, or finding the partition function of the corresponding classical model.

\section{Decimation Transform}
\label{sec:decimation}

In this section we consider a renormalization-group-like decimation transformation in the spirit of Ref.~\cite{LevinNave}, in order to analyze the effective long-distance physics. We show that models with crossing symmetry remain crossing symmetric under this transformation. We then show that crossing symmetry is {\em stable}: if the weights are varied slightly away from crossing symmetry, decimation causes the weights to ``flow'' back to those having crossing symmetry. We also show that these flows are toward ``infinite temperature'', where all Boltzmann weights are the same.

\begin{figure}[h]
\hspace*{80pt} \includegraphics[scale=0.15]{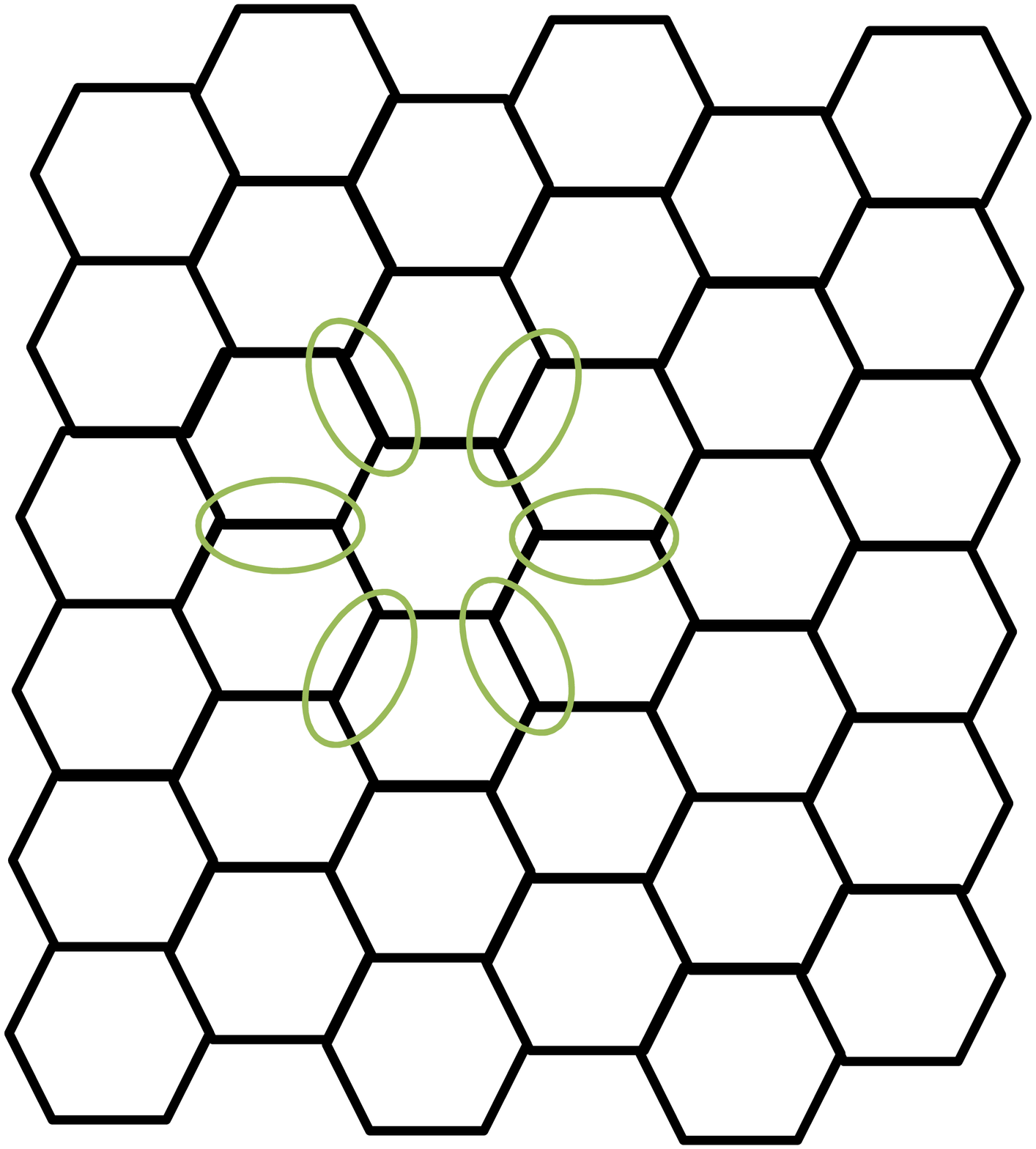}   ~~~~~
 \includegraphics[scale=0.15]{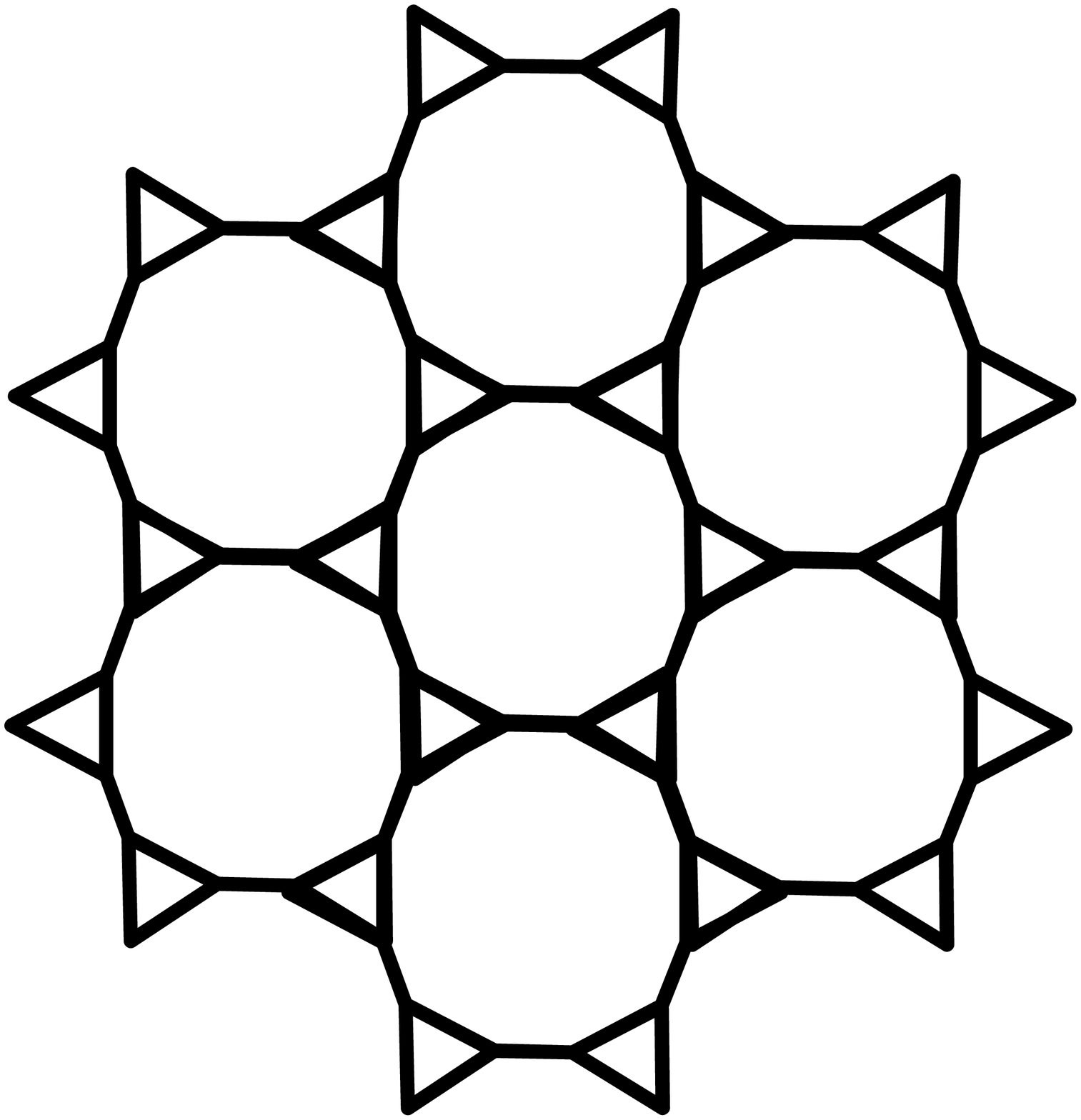}   ~~~~~
 \includegraphics[scale=0.15]{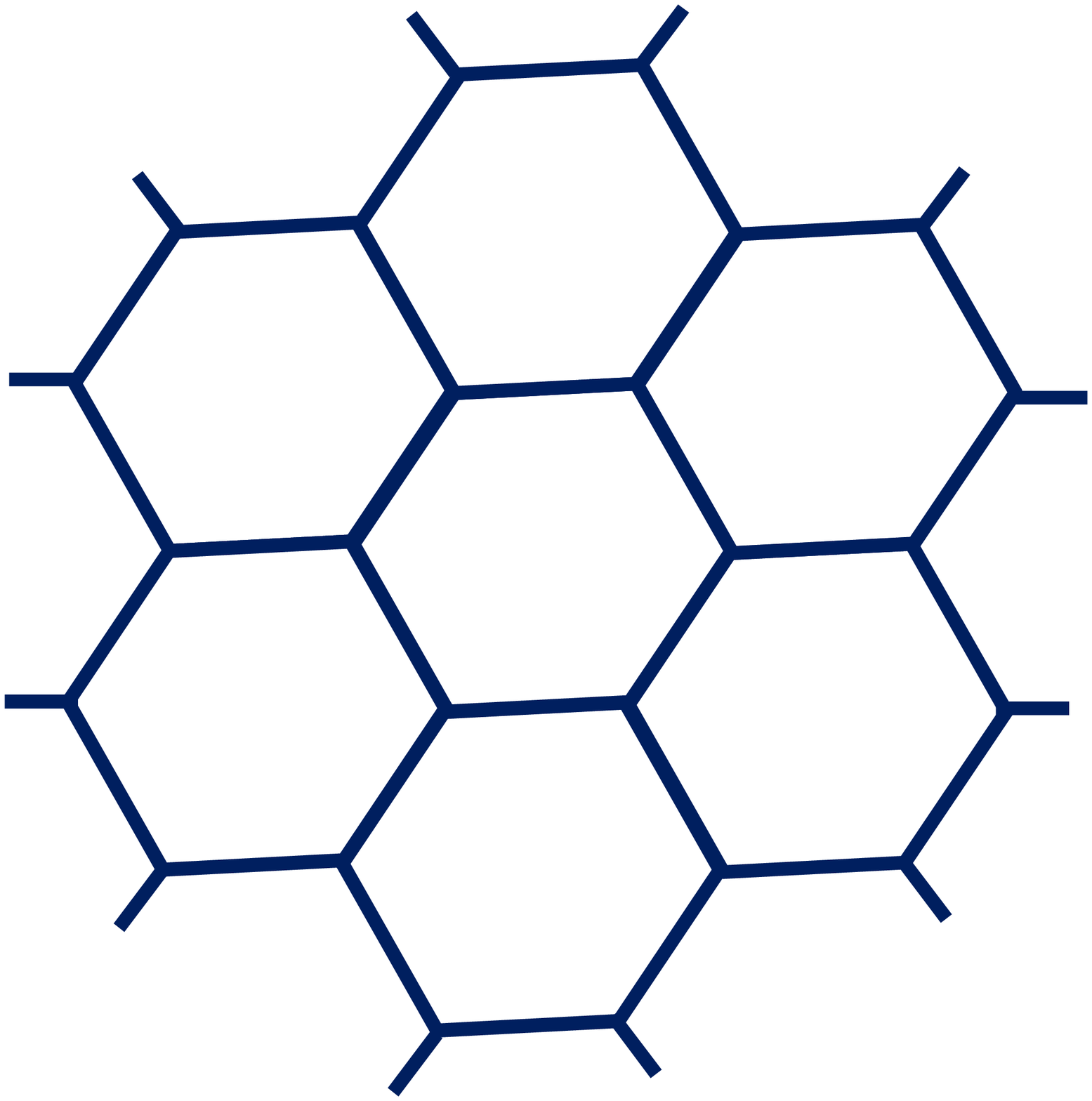}
 \caption{The steps of the RG transformation}
\label{fig:rg}
\end{figure}

\begin{figure}[h]
\hspace*{160pt}\begin{pspicture}(-1,0)(5,2)
\psline(-1,1.3)(-1,2)
\psline[arrowsize=5pt]{->}(-1,2)(-1,1.5)
\psline(-2,0)(-1.3,.7)
\psline[arrowsize=5pt]{->}(-2,0)(-1.5,.5)
\psline(0,0)(-.7,.7)
\psline[arrowsize=5pt]{->}(0,0)(-.5,.5)
\put(-.8,1.5){$a$}
\put(-2,.2){$b$}
\put(-.2,.2){$c$}
\psline(-1,1.3)(-1.3,.7)
\psline(-.7,.7)(-1.3,.7)
\psline(-1,1.3)(-.7,.7)
\put(1,1){$\rightarrow$}
\psline(3,1)(3,2)
\psline[arrowsize=5pt]{->}(3,2)(3,1.5)
\psline(2,0)(3,1)
\psline[arrowsize=5pt]{->}(2,0)(2.5,.5)
\psline(4,0)(3,1)
\psline[arrowsize=5pt]{->}(4,0)(3.5,.5)
\put(3.2,1.5){$a$}
\put(2,.2){$b$}
\put(3.8,.2){$c$}
\end{pspicture}
\caption{Reducing a decorated vertex to a vertex in the decimation}
\label{fig:reduce}
\end{figure}
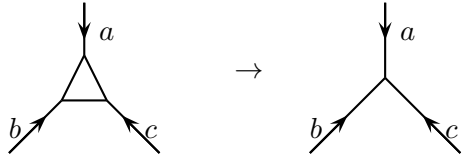

 The decimation transformation proceeds in steps as shown in Fig.~\ref{fig:rg}.
The transformation from the left to the middle proceeds via multiple F-moves (applied to the circled bonds, for example).   The transformation from the middle to the right requires turning a triangle (three vertices) into a single vertex via a transformation as in Fig.~\ref{fig:reduce}.
Mathematically, this transformation takes the form
$$
w(a,b,c) \rightarrow \sum_{i,j,k} w(a,i,j) w(b,k , \overline i) w(c, \overline j, \overline k)
$$
which constitutes the sum over the degrees of freedom of the triangle.

\subsection{The persistence of crossing symmetry}
\label{sub:someexact}

Crossing symmetry persists exactly under decimation.  Namely, after the decimation transformation is applied to a crossing-symmetric set of weights, the resulting set of weights will also satisfies crossing symmetry. This is easy to show by drawing pictures:
\begin{figure}[h]
\hspace*{80pt}{\psset{unit=.6cm}
\begin{pspicture}(-1,0)(5,2)
\put(-.5,1){$a$}
\put(-.5,-1){$d$}
\put(4.2,-1){$c$}
\put(4.2,1){$b$}
\psline(0,1)(1,0)
\psline(.5,.5)(.5,-.5)
\psline(0,-1)(1,0)
\psline(4,-1)(3,0)
\psline(4,1)(3,0)
\psline(3,0)(1,0)
\psline(3.5,.5)(3.5,-.5)
\put(4.6,0){$=$}
\put(5.5,1){$a$}
\put(5.5,-1){$d$}
\put(10.2,-1){$c$}
\put(10.2,1){$b$}
\psline(6,1)(7,0)
\psline(6,-1)(7,0)
\psline(10,-1)(9,0)
\psline(10,1)(9,0)
\psline(9,0)(7,0)
\pscircle[fillcolor=white,linecolor=black,fillstyle=solid](6.5,.5){.25}
\pscircle[fillcolor=white,linecolor=black,fillstyle=solid](9.5,-.5){.25}
\put(11,0){$=$}
\psline(13,1)(12,2)
\psline(14,2)(13,1)
\psline(13,1)(13,-1)
\psline(13,-1)(12,-2)
\psline(14,-2)(13,-1)
\put(11.5,2){$a$}
\put(11.5,-2){$d$}
\put(14.5,-2){$c$}
\put(14.5,2){$b$}
\pscircle[fillcolor=white,linecolor=black,fillstyle=solid](12.5,1.5){.25}
\pscircle[fillcolor=white,linecolor=black,fillstyle=solid](13.5,-1.5){.25}

\put(15,0){$=$}
\psline(17,1)(16,2)
\psline(18,2)(17,1)
\psline(17,1)(17,-1)
\psline(17,-1)(16,-2)
\psline(18,-2)(17,-1)
\put(15.5,2){$a$}
\put(15.5,-2){$d$}
\put(18.5,-2){$c$}
\put(18.5,2){$b$}
\psline(16.5,1.5)(17.5,1.5)
\psline(16.5,-1.5)(17.5,-1.5)

\end{pspicture}
}
\end{figure}
\vspace*{30pt}

This is nicely illustrated by the two-state models. For the $M=2$ vertex model described in section \ref{sec:vertex}, the decimation transformation is
\begin{align*}
\alpha & \rightarrow 2 \beta^3 + 6 \alpha \beta \tilde \beta  \\
\tilde \alpha & \rightarrow 2 \tilde \beta^3 + 6 \tilde \alpha \beta \tilde \beta  \\
\beta & \rightarrow 2 \alpha \tilde \beta^2 + 2 \alpha \tilde \alpha \beta  + 4 \beta^2 \tilde \beta \\
\tilde \beta & \rightarrow 2 \tilde \alpha \beta^2 + 2 \alpha \tilde \alpha \tilde \beta + 4 \beta \tilde \beta^2
\end{align*}
These still satisfy the crossing constraint  $\alpha \tilde \alpha = \beta \tilde \beta$ if the original weights do. To simplify the resulting weights, we can use the rescaling transformation (\ref{eq:rescale}) to rescale $\tilde \beta$ to unity.  Furthermore, with crossing symmetry we can always eliminate $\tilde \alpha$ using the constraint.  These allow us to obtain the simplified decimation transformation
\begin{align*}
\eta  & \rightarrow \frac{(1+3 \eta)^2}{\eta (3 + \eta)^2} \\
\beta & \rightarrow 4 \beta^3  (1 + 3 \eta) (3+ \eta) / \eta
\end{align*}
where we have defined
$ \eta= \beta^2/\alpha.$
The $\eta$ flow has only a single nonnegative attractive fixed point at $\eta=1$.  Using the rescaling (\ref{eq:rescale}) again, it is then easy to establish that this fixed point corresponds to a situation where all four types of vertex have equal weight.  This is a sensible outcome: if we rescale a huge number of vertices into a single vertex, the total partition function is almost entirely independent of the labels on the incoming bonds. One can then think of this as a flow to infinite temperature, where all Boltzmann weights are identical.

For the $M=2$ unoriented model shown in Fig.~\ref{fig:weightings}, the decimation transformation is
\begin{align*}
\alpha & \rightarrow \alpha^3 + 3 \beta^2 (\alpha + \gamma) + \gamma^3\\
\delta & \rightarrow \delta^3 + 3 \gamma^2 (\delta + \beta) + \beta^3 \\
\beta & \rightarrow \alpha^2 \beta + \delta \beta^2 + \beta^3 + 2 \alpha \beta \gamma + \delta \gamma^2 + 2 \beta \gamma^2 \\
\gamma & \rightarrow \alpha \beta^2 + \delta^2 \gamma + 2 \delta \beta \gamma + 2 \beta^2 \gamma + \alpha \gamma^2 + \gamma^3
\end{align*}
This indeed preserves the crossing condition
$
 \alpha \gamma  + \beta \delta = \beta^2 + \gamma^2
$.
Except for a few singular starting points where certain weights are zero, the system will always flow to the following form under decimation
$$
 (\alpha, \beta, \gamma, \delta) = K ( 1, x, x^2, x^3)
$$
with $K$ and $x$ some constants.  This form is stable under decimation, with
$$
 K \rightarrow K^3 ( 1 + x^2)^3
$$
for one step of the decimation transformation.  Comparing to Fig.~\ref{fig:weightings}, we see that each ``up" spin has fugacity $x$ and $K$ is the remaining weight of the vertex.

\subsection{Flows to crossing symmetry}
\label{sub:flowstocrossing}

We show here that crossing symmetry is stable under this decimation transformation. Namely, a set of weights that almost satisfies crossing symmetry flows {\em toward} crossing symmetry under decimation.  Our analysis is
very much in the spirit of a real-space renormalization-group procedure.

Let us consider Boltzmann weights $w = w_c + \epsilon \delta w$, where $w_c$ is a set of weights having crossing symmetry, $\delta w$ is a deviation from these values and $\epsilon \ll 1$ is a small parameter.     The partition function is written as
\be
 Z ({\cal G})= \sum_{\mbox{edge labels}}\,  \prod_{\mbox{vertices $v$}} \left[ w_c(j_v,k_v,l_v) +  \epsilon \delta w(j_v,k_v,l_v) \right]
\ee
Now let us expand this partition function order by order in $\epsilon$. The first two terms are
\begin{eqnarray*}
 Z ({\cal G})&=& \left( \sum_{\mbox{edge labels}}\,  \prod_{\mbox{vertices $v$}} w_c(j_v,k_v,l_v) \right)   \\
    &+& \epsilon \left( \sum_{\mbox{edge labels}}\, \sum_{\mbox{vertex $V$}} \delta w(j_v,k_v,l_v)   \prod_{\mbox{vertices $v \neq V$}} w_c(j_v,k_v,l_v) \right) + \ldots
\end{eqnarray*}
The first term is the partition function for the model with the crossing symmetric weights $w_c$. The order $\epsilon$ term is a sum over terms, where in each term a single vertex is given the perturbation $\delta w$ and all other vertices are given the crossing symmetric weights $w_c$.

It is useful to introduce $Z_c({\cal G}; i_{a_1}, i_{a_2}, i_{a_3}, \ldots, i_{a_p})$, which is the partition function given by not summing over the labels on the edges $a_1, \ldots, a_p$; on these edges the states are fixed to be $i_{a_1}, i_{a_2}, i_{a_3}, \ldots, i_{a_p}$.
We thus can write the full partition function as
$$
Z = Z_c({\cal G}) + \epsilon \sum_V  \sum_{j_V,k_V,l_V} \delta w(j_V,k_V,l_V)   Z_c({\cal G}; j_V,k_V,l_V) + \ldots
$$
where $j_V,k_V,l_V$ are the values of the edges incident on vertex $V$.   There is a slight complication if the graph has a ``tadpole''; at the tadpole vertex $j_V, k_V, l_V$ are not all independent. Here we have a sum over only two edge variables, say $j_V, k_V$, the weight would be $\delta w(j_V,k_V,k_V)$, and correspondingly we would have $Z_c({\cal G}; j_V,k_V)$. The next term in the series,  of order $\epsilon^2$, involves sums over two vertices $V_1$ and $V_2$. It will generically have a sum over the six edges connected to them. except when the two vertices are adjacent to each other there will again be fewer independent sums.
Typically the partially summed partition function $Z_c({\cal G}; i_{a_1}, i_{a_2}, i_{a_3}, \ldots, i_{a_p})$ is independent of geometry, because F-moves on edges other than $a_1,\dots a_p$ leave it invariant. Thus it is independent of the shape of $\cal G$ except in situations such as having three fixed edge labels $i_1, i_2, i_3$ all intersect at the same vertex; they cannot be geometrically rearranged by F-moves.

\hspace*{3cm} \begin{figure}[h]
 ~~~~~\includegraphics[scale=0.2]{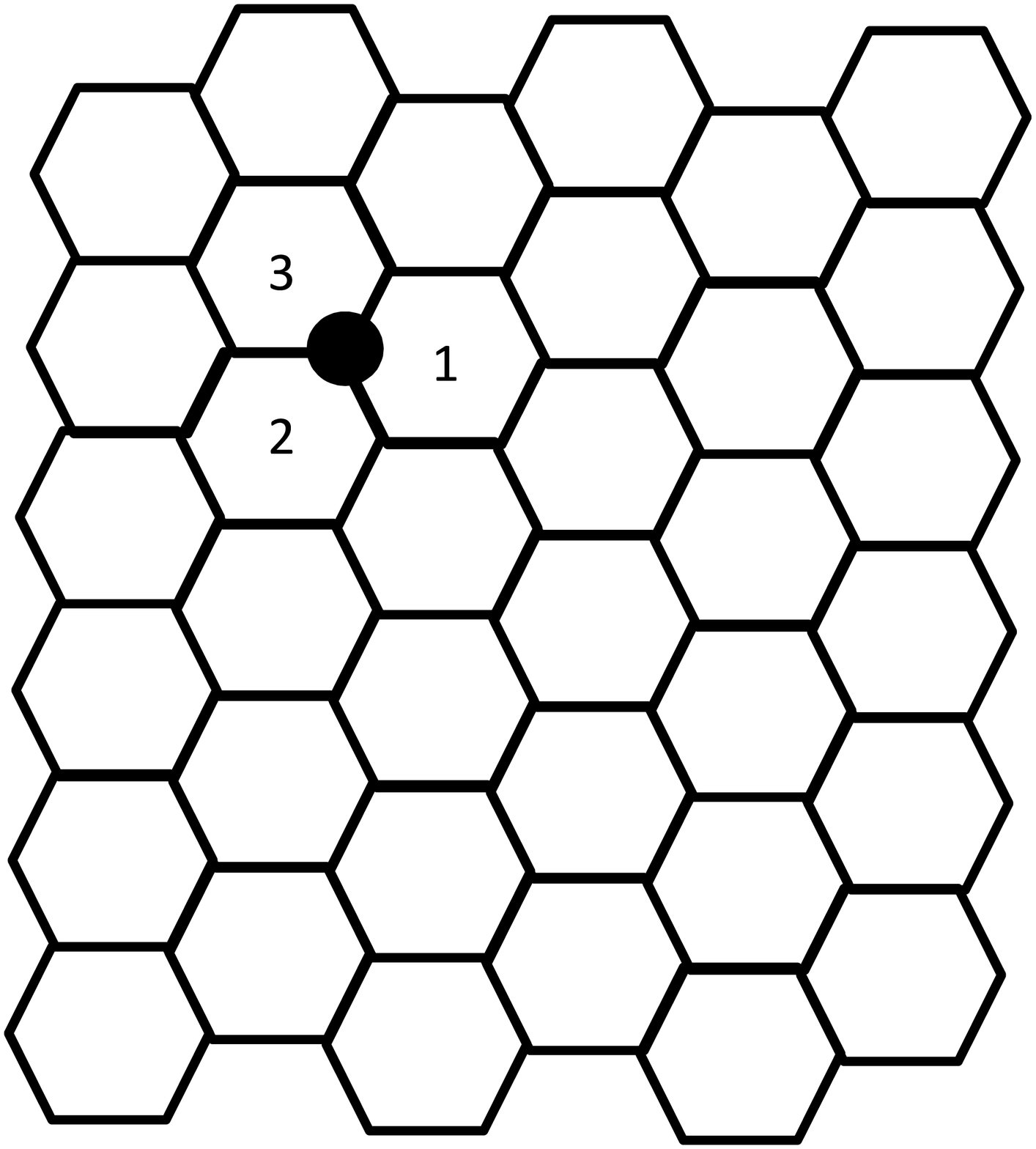}   ~~
 \includegraphics[scale=0.2]{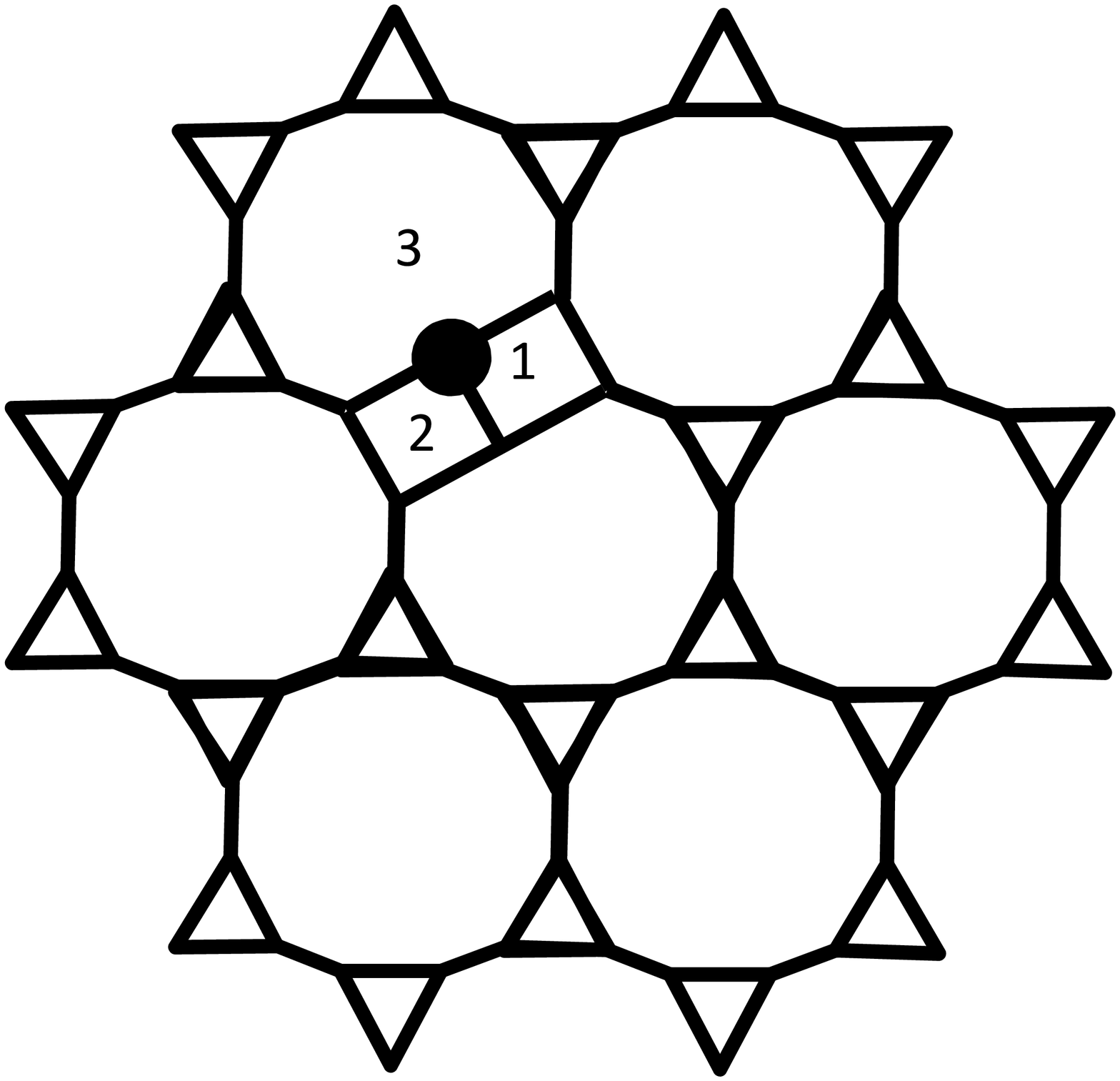}  ~~
\raisebox{-10pt}{\includegraphics[scale=0.2]{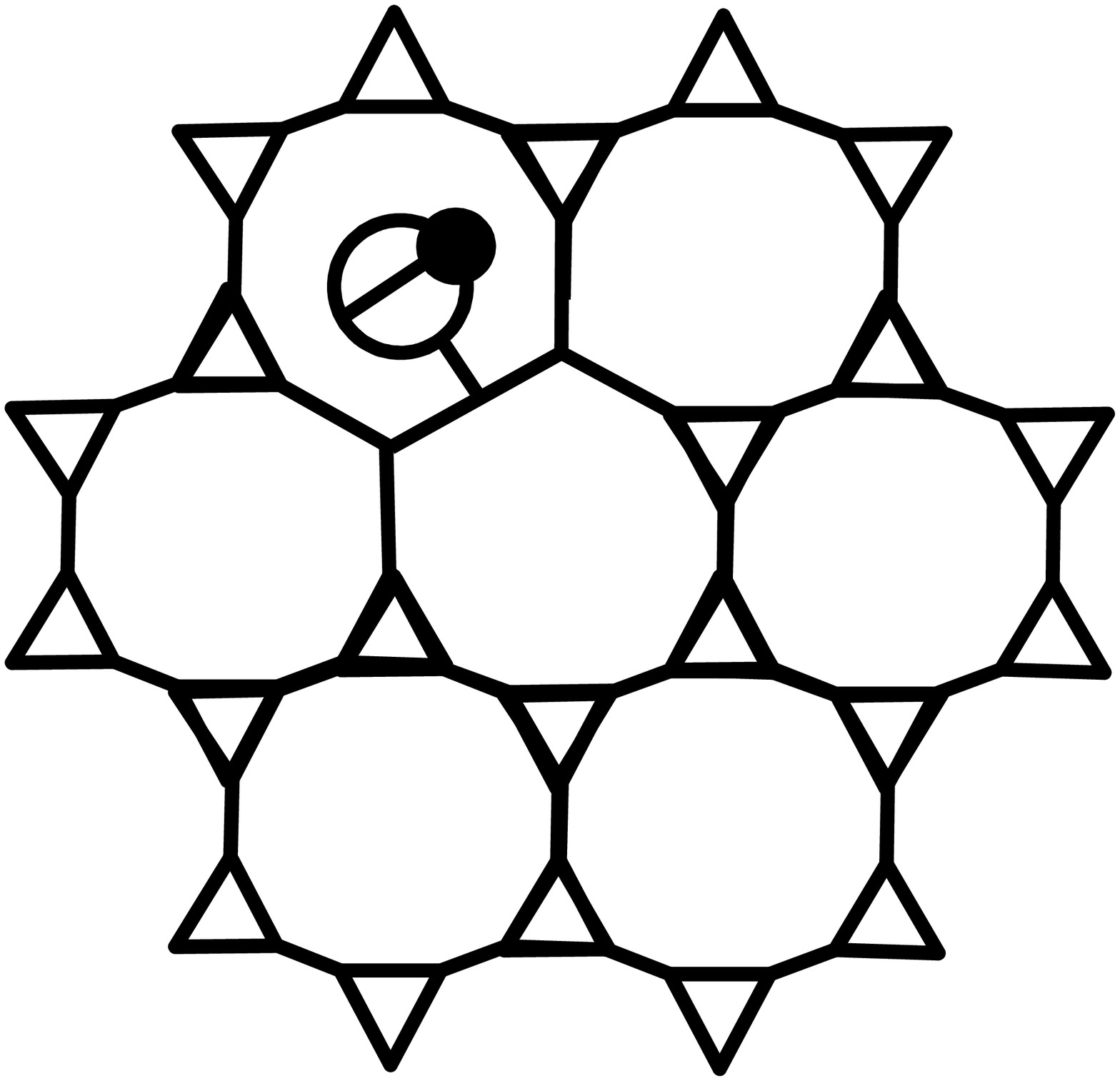}}  ~~
 \caption{The steps of the decimation transformation with one marked vertex.  In the left two panels, the cells adjacent to this vertex are numbered for clarity.}
\label{fig:rg2}
\end{figure}

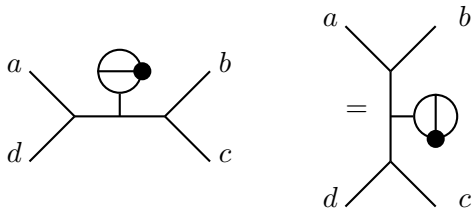
\begin{figure}[h]
\hspace*{50pt}{\psset{unit=.6cm}
\begin{pspicture}(-1,0)(5,2)
\put(-.5,1){$a$}
\put(-.5,-1){$d$}
\put(4.2,-1){$c$}
\put(4.2,1){$b$}
\psline(0,1)(1,0)
\psline(0,-1)(1,0)
\psline(4,-1)(3,0)
\psline(4,1)(3,0)
\psline(3,0)(1,0)
\psline(2,0)(2,.5)
\pscircle(2,1){.5}
\psline(1.5,1)(2.5,1)
\pscircle[fillcolor=black,fillstyle=solid](2.5,1){.2}
\put(7,0){$=$}
\psline(8,1)(7,2)
\psline(9,2)(8,1)
\psline(8,1)(8,-1)
\psline(8,-1)(7,-2)
\psline(9,-2)(8,-1)
\put(6.5,2){$a$}
\put(6.5,-2){$d$}
\put(9.5,-2){$c$}
\put(9.5,2){$b$}
\psline(8,0)(8.5,0)
\pscircle(9,0){.5}
\psline(9,.5)(9,-.5)
\pscircle[fillcolor=black,fillstyle=solid](9,-.5){.2}
\end{pspicture}
}
\vspace*{40pt}
\caption{F-moves still work when an edge is decorated with a tadpole having a marked vertex}
\label{fig:renF}
\end{figure}

Diagrammatically, let us indicate that a particular vertex takes the value $\delta w$ rather than $w_c$ by marking it with a dot, as shown on the left of Fig.~\ref{fig:rg2}.  To lowest order in $\epsilon$ we need only sum over diagrams each with a single marked vertex.   Now consider performing the above described decimation transformation.  Since the marked vertex does not obey crossing symmetry, we cannot perform an F-move on it and we obtain the figure in the middle of Fig.~\ref{fig:rg2}.   Let us focus in on the region around the marked dot.  We can still use F-moves around this region to  then transform to the right of Fig.~\ref{fig:rg2}.   Near the marked point,  note in particular that one can still use F-moves around this object as shown in Fig.~\ref{fig:renF}

\noindent  In the last step of the decimation, we would like to  to reduce the decorated vertices to single points, as shown in Fig.~\ref{fig:reduce}.  However, recall that the vertices attached to the marked point are not decorated at all.  Note that using F-moves we can rewrite the decorated vertex as

\hspace*{80pt}\begin{pspicture}(-1,0)(5,2)
\psline(-1,1.3)(-1,2)
\psline(-2,0)(-1.3,.7)
\psline(0,0)(-.7,.7)
\psline(-1,1.3)(-1.3,.7)
\psline(-.7,.7)(-1.3,.7)
\psline(-1,1.3)(-.7,.7)
\put(1,1){$=$}
\psline(3,1)(3,1.2)
\psline(3,1.6)(3,2)
\pscircle(3,1.4){.2}
\psline(2,0)(3,1)
\psline(4,0)(3,1)
\end{pspicture}

\noindent
Equivalently, the bubble can be placed on either of the other two legs.  We recall that the bubble is precisely the transfer matrix $T$ defined above.   Let us then define the matrix $R = T^{-1}$, and notate $R$ as a box.  Consider now the decorated vertex
\vspace*{20pt}

\begin{pspicture}(-1,0)(5,2)
\psline(3,1)(3,2.2)
\psframe[fillcolor=black,fillstyle=solid](2.85,1.65)(3.15,1.95)
\psline(3,1.2)(3.5,1.2)
\pscircle(3.9,1.2){.4}
\psline(3.9,1.6)(3.9,.8)
\pscircle[fillcolor=black,fillstyle=solid](3.9,.8){.15}
\psline(2,0)(3,1)
\psline(4,0)(3,1)
\end{pspicture}

\noindent
This vertex when connected to a vertex decorated with a triangle (like the left of Fig. \ref{fig:reduce}) removes the triangular decoration and leaves us with only the tadpole.  In a picture,

{\psset{unit=.6cm}
\begin{pspicture}(-1,0)(5,5)
\psline(2,4.5)(2.5,4)
\psline(4,4.5)(3.5,4)
\psline(3.5,4)(2.5,4)
\psline(3.5,4)(3,3)
\psline(2.5,4)(3,3)

\psline(3,1)(3,3)
\psframe[fillcolor=black,fillstyle=solid](2.85,1.65)(3.15,1.95)
\psline(3,1.2)(3.5,1.2)
\pscircle(3.9,1.2){.4}
\psline(3.9,1.6)(3.9,.8)
\pscircle[fillcolor=black,fillstyle=solid](3.9,.8){.15}
\psline(2,0)(3,1)
\psline(4,0)(3,1)

\put(7,2){$=$}
\psline(8,3)(7,4)
\psline(9,4)(8,3)
\psline(8,3)(8,1)
\psline(8,1)(7,0)
\psline(9,0)(8,1)
\psline(8,2)(8.5,2)
\pscircle(9,2){.5}
\psline(9,2.5)(9,1.5)
\pscircle[fillcolor=black,fillstyle=solid](9,1.5){.2}
\end{pspicture}}
\vspace*{10pt}

\begin{figure}
\hspace*{50pt}\begin{pspicture}(-1,0)(5,2)
\psline(-1,1.3)(-1,2)
\psline(-2,0)(-1.3,.7)
\psline(0,0)(-.7,.7)
\psline(-1,1.3)(-1.3,.7)
\psline(-.7,.7)(-1.3,.7)
\psline(-1,1.3)(-.7,.7)

\put(1,1){$+$}

\psline(3,1)(3,1.2)
\psline(3,1.6)(3,2)
\psline(2,0)(3,1)
\psline(4,0)(3,1)
\psline(3,1)(3,2.2)
\psframe[fillcolor=black,fillstyle=solid](2.85,1.65)(3.15,1.95)
\psline(3,1.2)(3.5,1.2)
\pscircle(3.9,1.2){.4}
\psline(3.9,1.6)(3.9,.8)
\pscircle[fillcolor=black,fillstyle=solid](3.9,.8){.15}
\psline(2,0)(3,1)
\psline(4,0)(3,1)

\put(5,1){$+$}

\psline(7,1)(7,1.2)
\psline(7,1.6)(7,2)
\psline(6,0)(7,1)
\psline(8,0)(7,1)
\psline(7,1)(7,2.2)
\rput*{45}(6.5,.5){
\psframe[fillcolor=black,fillstyle=solid](-.15,-.15)(.15,.15)}

\rput*{135}(9.72,-.5){
\psline(3,1.2)(3.5,1.2)
\pscircle(3.9,1.2){.4}
\psline(3.9,1.6)(3.9,.8)
\pscircle[fillcolor=black,fillstyle=solid](3.9,.8){.15}
}

\put(9,1){$+$}

\psline(11,1)(11,1.2)
\psline(11,1.6)(11,2)
\psline(10,0)(11,1)
\psline(12,0)(11,1)
\psline(11,1)(11,2.2)
\rput*{45}(11.5,.5){
\psframe[fillcolor=black,fillstyle=solid](-.15,-.15)(.15,.15)}

\rput*{45}(10,-2.28){
\psline(3,1.2)(3.5,1.2)
\pscircle(3.9,1.2){.4}
\psline(3.9,1.6)(3.9,.8)
\pscircle[fillcolor=black,fillstyle=solid](3.9,.8){.15}
}
\end{pspicture}
\caption{Four objects to be renormalized into a single vertex}
\label{fig:renv2}
\end{figure}
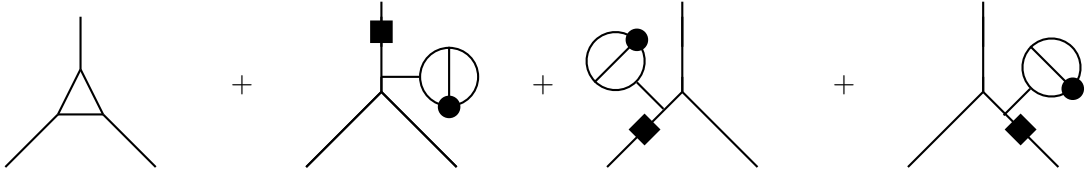

\noindent
Recall that we would like to add up all diagrams of the form of the right of Fig.~\ref{fig:rg2} with all possible bonds marked.   This can be done by using a renormalized vertex where the renormalized vertex is of the form of Fig.\ref{fig:renv2}.

\vspace*{20pt}

We have thus found the decimation transformation explicitly.   Each new vertex is comprised of the four diagrams in Fig.~\ref{fig:renv2}.   Building a hexagonal lattice out of these renormalized vertices precisely reproduces the partition function (and the free energy!) to order $\epsilon$.   What is remarkable is that, due to the relation shown in Fig.~\ref{fig:renF}, the new vertices obey the crossing relation to order $\epsilon$--- even though the original ones did not! Were we to consider perturbations to order $\epsilon^2$ we would require two iterations of the decimation to bring the system to crossing symmetry.

Thus we conclude that models perturbed slightly from crossing symmetry flow to crossing symmetry under decimation, in the fashion of a real-space renormalization-group flow.  This is natural when coupled with the observation that crossing-symmetric models themselves flow to having all Boltzmann weights equal. Thus in all cases the flow is toward infinite temperature. Models with crossing symmetry therefore are in a ``high-temperature'' phase.

\section{Summary and Outlook}
\label{sec:summary}

We have shown here how crossing symmetry may be used to solve certain classical statistical models with ease.  Several of these give solutions to simple-to-pose combinatorial problems, such as the number of closed nets on any trivalent graph. We described methods of finding more solutions to the crossing constraints, indicating a deep connection with the fusion algebra of anyon theories. We also argued by using a decimation transformation that such models behave as if they are in a high-temperature phase.

Several key tasks are left for further research.  We have explored only a few of the models with crossing symmetry; presumably there are other families of particularly interesting special cases. 
It also might be amusing to study these models on higher-genus or higher-dimensional surfaces, which are non-trivial to understand for oriented vertices. 
A better understanding of the appropriate mathematical classification of  models with crossing symmetry remains an interesting yet difficult task.  It also would be interesting to explore if useful quantum models other than the string-net models could be constructed from these classical models.  Finally, it would be interesting to explore whether there are quantum theories where Feynman diagrams may be resummed using the same type of crossing symmetries exploited here.

\medskip
We would like to thank Jesper Jacobsen for useful comments.  S.H.S.\ would like to thank Fiona Burnell, Curt von Keyserlingk, and Karin Erdmann for helpful conversations.   The work of P.F.\ is supported by the U.S. National Science Foundation under the grant DMR/MPS1006549, while the work of S.H.S.\ is supported by EPSRC Grant Numbers EP/I032487/1 and EP/I031014/1.

\appendix

\section{Three-state models ($M=3$)}
\label{app:m3}

For three states per links,  there are ten possible weights at each vertex. We label these as
$$\begin{array}{c}
 w(000) = \alpha,\qquad
 w(001) = \beta,\qquad
 w(002) = \gamma,\qquad
 w(011) = \delta\qquad
 w(022) = \epsilon,\\
 w(012) = \phi,\qquad
 w(111) = \eta,\qquad
 w(222) = \xi,\qquad
 w(122) = \kappa,\qquad
 w(211) = \mu.
 \end{array}
$$
We can also allow the weight to depend on the orientation of the vertex, so that  $w(012)=\phi$ is not necessarily the same as that of
$w(021)\equiv\tilde \phi$.  However, the crossing symmetry of the three-state model then requires
$$
\alpha(\phi-\tilde{\phi}) =
\beta(\phi-\tilde{\phi}) = \gamma(\phi-\tilde{\phi}) = \delta(\phi-\tilde{\phi}) = \epsilon(\phi-\tilde{\phi}) = \eta(\phi-\tilde{\phi}) = \xi(\phi-\tilde{\phi}) = \kappa(\phi-\tilde{\phi}) =\mu(\phi-\tilde{\phi})= 0\
$$
which then requires $\phi=\tilde \phi$. This is true independent of whether we are discussing the unoriented (self-conjugate) model described in \ref{app:unoriented} or the vertex (oriented) model described in \ref{app:vertex}.

\subsection{Unoriented edges}
\label{app:unoriented}

Consider the case where edges are all unoriented, i.e., they are all self-conjugate and we draw the edge labels without arrows.
Given the weights described above and the constraint $\phi=\tilde \phi$ there are then six remaining independent crossing constraints
$$
\begin{array}{l}
 -\phi^2-\kappa^2-\mu^2+\delta \epsilon+\eta \kappa+\xi \mu = 0 \\
 \beta \epsilon-\kappa \epsilon-\gamma \phi+\phi \xi+\delta \kappa-\phi \mu  = 0 \\
 -\gamma \delta+\mu \delta+\beta \phi-\phi \eta+\phi \kappa-\epsilon \mu = 0 \\
 \beta^2-\eta \beta+\delta^2+\phi^2-\alpha \delta-\gamma \mu  = 0 \\
 \beta \gamma-\kappa \gamma-\alpha \phi+\delta \phi+\epsilon \phi-\beta \mu  = 0 \\
 \gamma^2-\xi \gamma+\epsilon^2+\phi^2-\alpha \epsilon-\beta \kappa = 0
\end{array}
$$
and the transfer matrix is
$$
\left(
\begin{array}{lll}
 a^2+2 \beta^2+2 \gamma^2+\delta^2+\epsilon^2+2 \phi^2 & a \beta+2 \delta \beta+2 \gamma \phi+\delta \eta+\epsilon \kappa+2 \phi \mu & a \gamma+2 \epsilon \gamma+2 \beta \phi+\epsilon \xi+2 \phi \kappa+\delta \mu \\
 a \beta+2 \delta \beta+2 \gamma \phi+\delta \eta+\epsilon \kappa+2 \phi \mu & \beta^2+2 \delta^2+2 \phi^2+\eta^2+\kappa^2+2 \mu^2 & \beta \gamma+2 \delta \phi+2 \epsilon \phi+\xi \kappa+\eta \mu+2 \kappa \mu \\
 a \gamma+2 e \gamma+2 \beta \phi+\epsilon \xi+2 \phi \kappa+\delta \mu & \beta \gamma+2 \delta \phi+2 \epsilon \phi+\xi \kappa+\eta \mu+2 \kappa \mu & \gamma^2+2 \epsilon^2+2 \phi^2+\xi^2+2 \kappa^2+\mu^2
\end{array}
\right)
$$

\subsection{Vertex model}
\label{app:vertex}

Here we take two of the three states (let us choose 1,2) to be conjugate to each other, while state 0 is its own conjugate.  We can draw 1 and 2 with arrows and 0 as an edge without an arrow.  Given the weights described above and the constraint $\phi=\tilde \phi$, the remaining crossing constraints in the vertex model are
$$
\begin{array}{l}
 -\phi^2+\delta \epsilon+\eta \xi-\kappa \mu =0 \\
 \beta \epsilon-\mu \epsilon-\gamma \phi+\delta \xi  = 0 \\
 \phi^2-\alpha \phi+\beta \gamma+\delta \epsilon-\beta \kappa-\gamma \mu= 0 \\
 \beta^2-\mu \beta-\alpha \delta+2 \delta \phi-\gamma \eta =0 \\
 \gamma \delta-\kappa \delta-\beta \phi+\epsilon \eta = 0 \\
 \gamma^2-\kappa \gamma-\alpha \epsilon+2 \epsilon \phi-\beta \xi= 0 \\
\end{array}
$$
and the transfer matrix is
$$
\left(
\begin{array}{lll}
 \alpha^2+2 \phi^2+4 \beta \gamma+2 \delta \epsilon & \alpha \gamma+2 \phi \gamma+2 \beta \epsilon+\delta \xi+2 \phi \kappa+\epsilon \mu & \alpha \beta+2 \phi \beta+2 \gamma \delta+\epsilon \eta+\delta \kappa+2 \phi \mu \\
 \alpha \beta+2 \phi \beta+2 \gamma \delta+\epsilon h+\delta \kappa+2 \phi \mu & 2 \phi^2+\beta \gamma+2 \delta \epsilon+\eta \xi+3 \kappa \mu & \beta^2+2 \mu^2+4 \delta \phi+2 \eta \kappa \\
 \alpha \gamma+2 \phi \gamma+2 \beta \epsilon+\delta \xi+2 \phi \kappa+\epsilon \mu & \gamma^2+2 \kappa^2+4 \epsilon \phi+2 \xi \mu & 2 \phi^2+\beta \gamma+2 \delta \epsilon+\eta \xi+3 \kappa \mu
\end{array}
\right)
$$

\section{Some facts about the loop models}

\subsection{Proving the truncation $\Phi^{(M)}=0$}
\label{app:trunc}

Here we show that for $M$-dimensional matrices $\Phi^{(M)}=0$. Let us define the matrix
$$
  Y^{(j)} = (-1)^j \left[ \prod_{i=0}^{j-1} \alpha_i \right] \Phi^{(j)}
$$
The recursion relation (Eq. \ref{recursion}) for $Y$ is given by
\be
\label{recy}
  Y^{(j+1)} = -\Phi^{(1)} Y^{(j)} - \alpha_{j-1}^2 Y^{(j-1)}
\ee
and we want to show $Y^{(M)} =0$.

Note that all the $Y$'s and $\Phi$'s have the same set of eigenvectors which we have labeled $v^{(n)}$.  Let us apply Eq.~\ref{recy} to one of these eigevectors, to obtain the recursion
\be
\label{eq:yjrec}
  y^{(j+1)}_n = -\phi_n y^{(j)}_n - \alpha_{j-1}^2 y^{(j-1)}_n
\ee
where $y^{(j)}_n$ is the eigenvalue of $Y^{(j)}$ corresponding to the eigenvector $v^{(n)}$ and   $\phi_n$ is the eigenvalue of $\Phi^{(1)}$ corresponding to this eigenvector.

Consider now computing the eigenvalues of $\Phi^{(1)}$, we can write the characteristic equation
\be
0 = \det \begin{pmatrix}
-\phi &\alpha_0&0&0&\dots&0 &0&0\\
\alpha_0&-\phi&\alpha_1&0&\dots &0&0&0 \\
0&\alpha_1&-\phi&\alpha_2&&0&0&0\\
\vdots&&&&&&&\vdots\\
0&0&0&0&\dots &\alpha_{M-3}&-\phi&\alpha_{M-2}\\
0&0&0&0&\dots&0&\alpha_{M-2}&-\phi
\end{pmatrix}
\ee
Define $X^{(n)}$ to be the truncation of this secular matrix to $n$ dimensions.  Thus we have
\begin{eqnarray*}
X^{(1)} &=& -\phi \\
X^{(2)} &=&
\begin{pmatrix}
-\phi &\alpha_0\\
\alpha_0&-\phi
\end{pmatrix} \\
X^{(3)} &=&
\begin{pmatrix}
-\phi &\alpha_0 & 0\\
\alpha_0&-\phi & \alpha_1 \\
0 & \alpha_1&-\phi
\end{pmatrix}
\end{eqnarray*}
and so forth.  Our characteristic equation is
\be
\label{chareq}
 \det X^{(M)} = 0
\ee
Now consider calculating the determinant of $X^{(j+1)}$ by method of minors.  We obtain the recursion relation
$$
 \det X^{(j+1)} = -\phi \det X^{(j)} - \alpha_{j-1}^2 \det X^{(j-1)}
$$
We recognize this as the same recursion relation as Eq.~\ref{eq:yjrec}.  Thus, if we choose a value of $\phi$ which satisfies Eq.~\ref{chareq} (i.e., for $\phi$ one of the eigenvalues $\phi_n)$, then we must also have $y_n^{(M)}=0$ as well.  Thus all the eigenvalues of $Y^{(M)}$ are zero, and $\Phi^{(M)}$ must be zero as well.

\subsection{Proving that the loop weights are symmetric}
\label{app:symm}

Next we check that $w(j,k,l)=\Phi^{(j)}_{kl}$ is symmetric in all three indices, which is not very obvious from the definition. Since the matrices commute and $\Phi^{(0)}$ and $\Phi^{(1)}$ are symmetric, the recursion relation ensures that each matrix is symmetric, so that $w(j,k,l)=w(j,l,k)$. The remaining symmetry follows by explicitly plugging in the expression for $\Phi^{(1)}$ in the recursion relation, yielding
\begin{eqnarray}
\label{recur2}
\alpha_j\, \Phi^{(j+1)}_{k,l}+\alpha_{j-1}\Phi^{(j-1)}_{k,l}
&=&\alpha_{k}\,\Phi^{(j)}_{k+1,l} + \alpha_{k-1}\Phi^{(j)}_{k-1,l}\\
\label{recur3}
\alpha_j\, w(j+1,k,l)+\alpha_{j-1}\,w(j-1,k,l)&=&\alpha_{k}\,w(j,k+1,l) + \alpha_{k-1}\,w(j,k-1,l)
\ .
\end{eqnarray}
Consider instead a set of matrices $\Psi^{(k)}$ defined so that they have entries
$\Psi^{(j)}_{kl} =w(k,j,l)$. Then (\ref{recur3}) implies that
\[\alpha_j\, \Psi^{(k)}_{j+1,l}+\alpha_{j-1}\Psi^{(k)}_{j-1,l}
=\alpha_{k}\,\Psi^{(k+1)}_{j,l} + \alpha_{k-1}\Psi^{(k-1)}_{j,l}\ .
\]
Since this holds for all $j$ and $k$, this is the same recursion relation as (\ref{recur2}). This means if the matrices $\Psi^{(0)}=\Phi^{(0)}$ and $\Psi^{(1)}=\Phi^{(1)}$, this recursion requires that
$\Psi^{(j)}=\Phi^{(j)}$ for all $j$, and that the weights have the desired symmetry.
The last things to prove therefore are that  $w(j,0,k)=w(0,j,k)$ and $w(j,1,k)=w(1,j,k)$ For the first, assume that $w(m,0,k)=w(0,m,k)=\alpha_0\delta_{m,k}$ for all $m\le j$. Then (\ref{recur3}) with $l=0$ and the last two entries exchanged implies that
\[w(j+1,0,k)=\frac{\alpha_0}{\alpha_j}(-\alpha_{j-1}\delta_{j-1,k}+\alpha_{k}\delta_{j,k+1} + \alpha_{k-1}\delta_{j,k-1})=\alpha_0\delta_{j,k-1}=w(0,j+1,k)
\]
Since $w(1,0,k)=\alpha_0\delta_{k,1}$ by definition, recursion therefore implies $w(j,0,k)=w(0,j,k)$ for all $k$ as desired. The proof that $w(j,1,k)=w(1,j,k)$ follows analogously by setting $l=1$ in (\ref{recur3}).

\subsection{Pairing of Eigenvalues}
\label{app:pairs}

Consider the matrix $\Phi^{(1)}$.  Its eigenvalue equation is
$$
  \alpha_{k+1} v_{k+1} + \alpha_{k-1} v_{k-1} = \phi^{(1)} v_k
$$
with eigenvector $v_k$ with indices $k=0, \ldots M-1$ and eigenvalue $\phi^{(1)}$  (having defined here $\alpha_{-1} = \alpha_{M} =0$).    From the eigenvector $v_k$ we can construct another eigenvector $\tilde v_k$ given by
$$
  \tilde v_k = (-1)^k v_k
$$
which has eigenvalue $\tilde \phi^{(1)}=-\phi^{(1)}$.  Thus eigenvalues of $\Phi^{(1)}$ must come in pairs.  The exception to this is if $\phi^{(1)}=0$ which occurs if $M$ is odd which corresponds an eigenvector
with $v_k=0$ for all odd $k$.

Correspondingly let us define  $\phi^{(j)}$ to be the eigenvalue of $\Phi^{(j)}$ corresponding to the eigenvector $v_k$, and let  $\tilde \phi^{(j)}$ to be the eigenvalue of $\Phi^{(j)}$ corresponding to the eigenvector $\tilde v_k$. From the recursion Eq.~\ref{recursion} we see that these eigenvalues also come in pairs
$$
 \tilde \phi^{(j)}= (-1)^j \phi^{(j)}
$$
Thus using Eq.~\ref{eq:eigsT} we see that the eigenvalues of $T$ must also come in pairs, with the exception of one lone eigenvalue which occurs only in the case of $M$ odd, which corresponds to the eigenvector with zero eigenvalue for $\Phi^{(1)}$.  In the case of $M$ odd, we should ask whether the lone eigenvalue could possibly be the largest eigenvalue of $T$.  In fact, this can never occur by the Perron-Frobenius theorem, which guarantees that the eigenvector corresponding to the largest eigenvalue must have all positive entries.

\end{document}